\documentclass[letterpaper,twocolumn,10pt]{article}
\usepackage{usenix-2020-09} 

\usepackage{tikz}
\usepackage{amsmath}
\usepackage{amsfonts}
\usepackage{url}
\usepackage{filecontents}
\usepackage{xspace}
\usepackage{subfig}
\usepackage{multirow}
\usepackage{makecell}
\usepackage[ruled, linesnumbered]{algorithm2e}

\usepackage[T1]{fontenc}
\usepackage{inconsolata}

\usepackage{enumitem}

\usepackage{amssymb}
\usepackage{pifont}
\newcommand{\cmark}{\textcolor{black}{\ding{51}}}%
\newcommand{\xmark}{\textcolor{black}{\ding{55}}}%

\newcommand{\PHB}[1]{\vspace{.5em}\noindent\textbf{#1}} 
\newcommand{\SystemName}{\textsc{CaraServe}\xspace}
\newcommand*\circled[1]{\tikz[baseline=(char.base)]{\node[shape=circle,draw,inner sep=0.3pt] (char) {#1};}}

\def\huawei{{$^\dagger$}}
\def\shangailab{{$^\ddagger$}}
\def\cuhk{{$^\diamond$}}

\usepackage{authblk}

\begin{document}


\title{\Large \bf \SystemName: CPU-Assisted and Rank-Aware LoRA Serving for Generative LLM Inference}

\author{
{\rm Suyi Li$^*$, Hanfeng Lu$^*$, Tianyuan Wu, Minchen Yu\cuhk, Qizhen Weng\shangailab, Xusheng Chen\huawei, Yizhou Shan\huawei, Binhang Yuan, Wei Wang} \\
HKUST, {\cuhk}CUHK-Shenzhen, {\shangailab}Shanghai AI Laboratory, {\huawei}Huawei Cloud \\
} 

\maketitle

\def\thefootnote{*}\footnotetext{Equal contribution}

\begin{abstract}

Pre-trained large language models (LLMs) often need specialization for
domain-specific tasks. Low-Rank Adaptation (LoRA) is a popular approach that
adapts a base model to multiple tasks by adding lightweight trainable
adapters. In this paper, we present \SystemName, a system that efficiently
serves many LoRA adapters derived from a common base model. \SystemName
maintains the base model on GPUs and dynamically loads activated LoRA
adapters from main memory. As GPU loading results in a \emph{cold-start} that
substantially delays token generation, \SystemName employs a \emph
{CPU-assisted} approach. It early starts the activated adapters on CPUs for
prefilling as they are being loaded onto GPUs; after loading completes, it
then switches to the GPUs for generative LoRA inference. \SystemName develops
a highly optimized synchronization mechanism to efficiently coordinate LoRA
computation on the CPU and GPU. Moreover, \SystemName employs a \emph
{rank-aware} scheduling algorithm to optimally schedule heterogeneous LoRA
requests for maximum service-level objective (SLO) attainment. We have
implemented \SystemName and evaluated it against state-of-the-art LoRA
serving systems. Our results demonstrate that \SystemName can speed up the
average request serving latency by up to $1.4\times$ and achieve an SLO
attainment of up to 99\%.

\end{abstract}

\section{Introduction}
\label{sec:intro}

Large language models (LLMs) are making significant strides in generative
AI~\cite{llama2,zhang2022opt}, enabling a variety of novel applications across
numerous domains. Deploying LLMs for domain-specific tasks requires
specialization~\cite{openai_custom_instructions,qlora}, which involves
adapting a pre-trained base model to different downstream tasks. Low-Rank
Adaptation~\cite{hu2022lora,qlora,longlora} (LoRA) has emerged as a popular
parameter-efficient fine-tuning (PEFT) approach. It preserves the base
model's parameters and adds trainable rank decomposition matrices to each
Transformer layer. This method significantly reduces the number of trainable
parameters, allowing the creation of numerous lightweight LoRA adapters 
from a single base model. As LoRA gains popularity in LLM deployment, efficiently
serving them in a multi-tenant cloud becomes critically important~\cite
{chen2023punica,sheng2023slora}.

However, developing a system for efficient LoRA serving presents non-trivial
challenges. One straightforward solution is to \emph{merge} the weights of a
LoRA adapter into the parameters of the base model, resulting in an
independent, specialized LLM instance of a full size (e.g., HF-PEFT~\cite
{hf_peft}). This approach, though easy to implement, is expensive as it requires duplicating the
base model for individual LoRA instances, consuming a substantial amount of
GPU memory. Recently, pioneering attempts have been made to enable \emph
{base model multiplexing} between LoRA adapters~\cite
{chen2023punica,sheng2023slora}, in which the system maintains a shared copy
of the base LLM on the GPU and loads LoRA adapters from main memory
as requests arrive. Although this approach is GPU-efficient, it results in a
severe \emph{cold-start} problem when a requested LoRA adapter is not on GPU
and must be fetched from main memory. Depending on the adapter size, a single
cold-start can take tens of milliseconds. This delay affects not only the \emph
{time-to-first-token} of the newly arrived request but also the \emph
{decoding} process of other ongoing requests when \emph
{continuous batching}~\cite{yu2022orca, kwon2023vllm, lightllm, tgi_hf} is in
use, resulting in an average of 25\% latency increase in inference serving in our
experiments (\S\ref{subsec:challenges}).

\begin{table}[]
\centering
\footnotesize
\begin{tabular}{l|c|c|c}
\hline
                 & \textbf{GPU-Efficient} & \textbf{Cold-Start-Free} & \textbf{SLO-Aware} \\
\hline
HF-PEFT~\cite{hf_peft}          &   \xmark        &   \cmark         &  \xmark       \\
S-LoRA~\cite{sheng2023slora}            &   \cmark        &   \xmark         &  \xmark       \\
Punica~\cite{chen2023punica}           &   \cmark        &   \xmark         &  \xmark       \\
\SystemName      &   \cmark        &   \cmark         &  \cmark       \\
\hline
\end{tabular}
\caption{Summarization of LoRA serving systems.}
\label{tab:sys_compare}
\end{table}

\begin{figure}[t]
\centering
\includegraphics[width=0.9\linewidth]{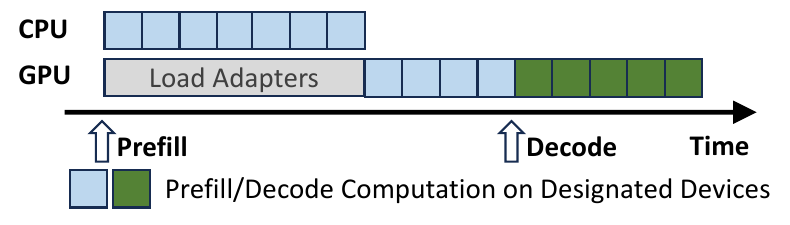}
\caption{Illustration of CPU-assisted LoRA serving.}
\label{fig:lora_flow_ite}
\end{figure}

We believe a desirable LoRA serving system should exploit base model
multiplexing for \emph{GPU-efficient inference}, without incurring high
cold-start overhead (\emph{cold-start-free}). Additionally, as a multi-tenant
system, it should prioritize meeting users' service-level objectives in
latency (\emph{SLO-aware}) by judiciously scheduling their inference requests
to heterogeneous LoRA models with varying ranks. Unfortunately, current
systems fail to fulfill these requirements (see the summarization in
Table \ref{tab:sys_compare}). To bridge this gap, we
present \SystemName (\underline{C}PU-\underline{a}ssisted, \underline{R}ank-\underline{a}ware Serve),
a multi-tenant LoRA serving system that achieves all
three design goals concurrently. We highlight the design approaches and key
techniques of \SystemName as follows:

\PHB{CPU-assisted LoRA serving.}
Similar to the existing LLM-multiplexing solutions~\cite
{chen2023punica, sheng2023slora}, \SystemName maintains the base LLM on GPUs
and all LoRA adapters in main memory, which are dynamically loaded onto the
GPU as new requests arrive. Yet, instead of waiting for the adapter loading to
complete, \SystemName \emph{concurrently} runs the adapter on CPU
to early-start the \emph{prefill} phase. 
Once the adapter is fully loaded, \SystemName switches to GPU computation to resume the prefill phase, if not finished, and then proceed to the subsequent decoding phase (Fig.~\ref{fig:lora_flow_ite}), alongside other ongoing requests using continuous batching~\cite{yu2022orca, kwon2023vllm, lightllm, tgi_hf}.
This \emph{CPU-assisted} approach effectively mitigates the
cold-start overhead, substantially improving decoding efficiency.

Nevertheless, implementing CPU-assisted LoRA serving poses several
challenges. LLMs are constructed using the Transformer~\cite
{vaswani2017attention} architecture, which consists of multiple attention layers. During inference, the computed output of the base LLM needs to be synchronized with that of the LoRA models at each layer. Since these computations are split between the CPU and GPU, 
efficient layer-wise synchronization between the two devices is crucial. Additionally, the frequent triggering of LoRA computations (e.g., 32 times per decoding iteration in Llama2-7B~\cite
{llama2}) leads to high invocation overheads, such as inter-process communication
(IPC) and data transfer, which can significantly increase inference latency by $79.4\%$. 
Moreover, offloading the heavy prefill computation to the CPU may create a new bottleneck due to
its limited parallelism compared with GPU.

We address these challenges with a series of techniques. To efficiently
coordinate on-GPU LLM computation and on-CPU LoRA computation, we develop a
specialized CUDA operator that optimally pipelines the two computations by
means of asynchronous memory copy and signaling. Additionally, we employ
shared memory to enable fast data exchange between the base LLM process and
multiple CPU LoRA processes, eliminating the need for data copying and
serialization. This reduces the LoRA invocation overhead to less than 1 ms.
Furthermore, we devise a profiling-guided parallelization scheme to scale out
LoRA computations across multiple CPUs to eliminate the potential bottleneck.
Putting it altogether, \SystemName can reduce the prefill latency by $57.9\%$.

\PHB{Rank-aware request scheduling.}
In multi-tenant LoRA serving, users often request to utilize heterogeneous
adapters with different ranks, which can be batched together to multiplex the
base LLM~\cite{chen2023punica,sheng2023slora}. However, we observe significant
performance variations in decoding when batching different sets of
heterogeneous LoRA adapters (\S\ref{subsec:challenges}). This highlights the
need for intelligent request scheduling that takes into account the rank
heterogeneity and its impact on decoding. To this end, we establish a
\emph{performance model} through extensive system profiling, which can be used to
accurately predict the decoding latency for a specific batch of LoRA adapters. 
Leveraging this information, we design a \emph
{rank-aware scheduling algorithm} to enhance cluster-wide performance and
meet users' latency SLOs. Specifically, when a new request arrives, the
scheduler evaluates all inference servers that possess the required LoRA adapters
and calculates a cost score for each server using the performance model. This score
measures the additional latency cost and SLO violation on the current
ongoing requests if the new request were to be accommodated in that server. 
The scheduler then selects the server with the minimum cost score and routes 
the request to it accordingly.

We have implemented \SystemName as a pluggable LLM serving module in
LightLLM~\cite{lightllm} and evaluated its performance using
Llama2-7B/30B/70B~\cite{llama2} with requests generated from synthetic and
real-world traces. Our evaluation highlights that \SystemName outperforms
S-LoRA~\cite{sheng2023slora}, the state-of-the-art solution, by accelerating
the average serving latency of inference requests by up to $1.4\times$.
We also evaluated the rank-aware scheduling algorithm through testbed
experiments and large-scale simulations. Compared to popular scheduling
policies, including the one used in the existing adapter serving system~\cite
{chen2023punica}, \SystemName reduces the average time per token by up to
36.4\% and achieves an SLO attainment of 99\%. 

We will release \SystemName as an open-source software after the double-blind
review process.

\newcommand{\x}{\mathbf{x}}
\newcommand{\xin}{\mathbf{x}_{\text{In}}}
\newcommand{\wq}{\mathbf{W}_Q}
\newcommand{\wv}{\mathbf{W}_V}
\newcommand{\wk}{\mathbf{W}_K}
\newcommand{\wo}{\mathbf{W}_O}
\newcommand{\wa}{\mathbf{W}_1}
\newcommand{\wb}{\mathbf{W}_2}
\newcommand{\xq}{\mathbf{x}_Q}
\newcommand{\xv}{\mathbf{x}_V}
\newcommand{\xk}{\mathbf{x}_K}
\newcommand{\xs}{\mathbf{x}_{\text{score}}}
\newcommand{\xsm}{\mathbf{x}_{\text{softmax}}}
\newcommand{\xa}{\mathbf{x}_{\text{Att}}}
\newcommand{\xo}{\mathbf{x}_{\text{Out}}}
\newcommand{\xma}{\mathbf{x}_{\text{MLP}_1}}
\newcommand{\xmb}{\mathbf{x}_{\text{MLP}_2}}

\newcommand{\tx}{\mathbf{t}}
\newcommand{\tq}{\mathbf{t}_Q}
\newcommand{\tv}{\mathbf{t}_V}
\newcommand{\tk}{\mathbf{t}_K}
\newcommand{\ts}{\mathbf{t}_{\text{score}}}
\newcommand{\tsm}{\mathbf{t}_{\text{softmax}}}
\newcommand{\ta}{\mathbf{t}_{\text{Att}}}
\newcommand{\txo}{\mathbf{t}_{\text{Out}}}
\newcommand{\tma}{\mathbf{t}_{\text{MLP}_1}}
\newcommand{\tmb}{\mathbf{t}_{\text{MLP}_2}}

\section{Background and Motivation}
\label{sec:background}

In this section, we give a primer to LLM inference and low-rank adaptation
(LoRA). We also discuss the key challenges that arise when serving
LoRA models in a multi-tenant cloud.

\subsection{LLM Inference}
\label{subsec:llm_inference}

\PHB{Generative LLM inference computation.} LLM inference is a process that involves generating a sequence of output tokens in response to an input prompt, which is a list of tokens. 
This process consists of two phases: \emph{prefill} and \emph{decoding}. During the prefill phase, the input sequence is used to generate the key-value cache (KV cache) for each transformer layer; the decoding phase then uses the previous KV cache to generate new tokens step-by-step and update the KV cache accordingly. The computation of one transformer layer can be summarized as follows.
Denote the batch size by $B$, the prompt sequence length by $L$, the hidden dimension of the transformer by $H$, and the intermediate size by $H'$. 
We have weight matrices of the $i$-th transformer layer: $\wk^i, \wq^i, \wv^i, \wo^i \in \mathbb{R}^{H \times H} $, $\wa^i \in \mathbb{R}^{H \times H'}$, and $\wb^i \in \mathbb{R}^{H' \times H}$.
During the prefill phase, let $\x^i$ be the input of the $i$-th transformer layer, and the \texttt{key}, \texttt{value}, \texttt{query}, and \texttt{output} of the attention layer respectively specified as $\xk^i, \xv^i, \xq^i, \xo^i \in \mathbb{R}^{B \times L \times H}$. The computation of
the cached \texttt{key}, \texttt{value} is given by
$\xk^i = \x^i \cdot \wk^i$ and $\xv^i = \x^i \cdot \wv^i$.
The remaining computation in this transformer layer is given by
\begin{equation*}
\small
\vspace{-.01in}
\begin{array}{cc}
& \xq^i = \x^i \cdot \wq^i, \\
& \xo^i = f_{\text{softmax}}\left( \frac{\xq^i {\xk^i}^T}{\sqrt{H}}\right )\cdot \xv^i \cdot \wo^i + \x^i, \\
& \x^{i+1} = f_{\text{relu}}\left(\xo^i \cdot \wa^i \right) \cdot \wb^i + \xo^i.
\end{array}
\vspace{-.01in}
\end{equation*}

During the decoding phase, let $\tx^i \in \mathbb{R}^{B \times 1 \times H}$ be the embedding of the current generated token in the $i$-th layer. The inference computation involves i) updating the KV cache, i.e., $\xk^i \leftarrow f_{\text{concat}}\left( \xk^i, \tx^i \cdot \wk^k \right)$, 
$\xv^i \leftarrow f_{\text{concat}}\left( \xv^i, \tx^i \cdot \wv^i \right)$,
and ii) computing the output of the current layer:
\begin{equation*}
\small
\begin{array}{cc}
     & \tq^i = \tx^i \cdot \wq^i, \\
     & \txo^i = f_{\text{softmax}}\left( \frac{\tq^i {\xk^i}^T}{\sqrt{H}}\right)\cdot \xv^i \cdot \wo^i + \tx^i,\\
     & \tx^{i+1} = f_{\text{relu}}\left(\txo^i\cdot \wa^i \right) \cdot \wb^i + \txo^i.
\end{array}
\vspace{-.02in}
\end{equation*}
The decoding phase continues until a specified condition is met, such as emitting
an end-of-sequence (\texttt{<eos>}) token or reaching a desired output sequence length.


\PHB{LLM adaption.}
\label{sec:llm_adaption}
Adapting LLMs in a parameter-efficient manner is a popular approach to enhancing
their performance for domain-specific tasks or customizing the model inference
results to align with human intents~\cite
{openai_custom_instructions, openai_gpt_finetuning}. One notable approach is
Low-Rank Adaptation or LoRA~\cite{hu2022lora}, which introduces an \emph{adapter} to
modify the intermediate LLM inference results while keeping the original LLM
parameters unchanged.
Specifically, given a pre-trained weight matrix $\mathbf{W} \in \mathbb{R}^{H_1 \times H_2}$, an adapter consists of two low-rank matrices $\mathbf{A} \in \mathbb{R}^{H_1 \times r}$ and $\mathbf{B} \in \mathbb{R}^{r \times H_2}$, where $r$ is the LoRA rank. LoRA adapts this weight matrix 
to $\mathbf{W}^{\prime} = \mathbf{W} + \mathbf{AB}$. 
Let $\mathbf{y}$ be the original output of this layer given by
$\mathbf{y} = \mathbf{x} \mathbf{W}$.
With LoRA adaption, the updated computation becomes
\begin{equation}
\small
  \label{eq:xW+xAB}
  \mathbf{y}^{\prime} = \mathbf{x} \mathbf{W} + \mathbf{x A B} = \mathbf{x W}^{\prime}.
\end{equation}

The LoRA adapter is highly efficient in terms of parameter space because the rank $r \times (H_1 + H_2) \ll H_1 \times H_2 $. Therefore, LoRA adaption is widely applied in the attention modules of transformer-based LLMs~\cite{hu2022lora, sheng2023slora}. When deploying LoRA-adapted models for inference, the computation load required by the LoRA adapter ($\mathbf{x A B}$) is orders of magnitude smaller than that of the original weights $\mathbf{x W}$ in terms of floating-point operations, if we compute these two parts separately.


\subsection{Multi-Tenant LoRA Serving}

\PHB{The need of LLM-multiplexing.}
A naive way to serve a LoRA adapter~\cite{hu2022lora} is to \emph{merge} its
weights into the weights of the base LLM, which introduces no additional
computational overhead when deploying the adapted model for inference. However,
this approach does not scale to multi-tenant LoRA serving: because one base
model can only merge with one LoRA adapter at a time, serving $n$ different LoRA
models requires duplicating $n$ copies of the base LLM, wasting GPU memory and 
missing opportunities for batch inference~\cite{kwon2023vllm}.

In practice, many LoRA models are developed based on common LLM series (e.g., Llama2~\cite
{llama2}), and multiple LoRA models originating from the same LLM can
multiplex that LLM for GPU-efficient inference. This can be
achieved by computing LoRA adaption $\mathbf{xAB}$ on the fly and adding this
result back to the intermediate results $\mathbf{xW}$ before subsequent computations. As described in \S\ref
{sec:llm_adaption}, the computation of $\mathbf{xAB}$ is lightweight, and
multiple LoRA computations can be batched during inference.

\begin{figure}[tbp]
    \centering
    \includegraphics[width=0.95\linewidth]{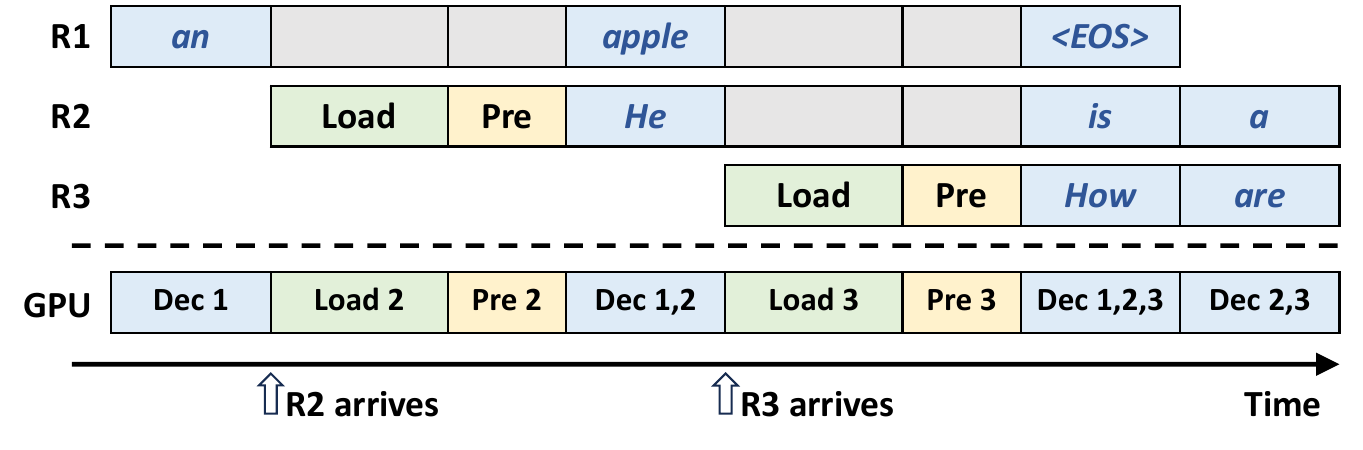}
    \caption{Continuous batching in which the decoding phase (Dec) is preempted to perform prompt processing upon a request arrival, which involves loading the requested LoRA adapter
    (Load) and prefilling (Pre).}
    \label{fig:continuous_batching}
    \vspace{-.05in}
\end{figure}

\PHB{Continuous batching.}
\label{sec:continuous_batching}
Existing LLM serving systems employ a \emph{continuous batching} strategy 
optimized for LLM's iterative auto-regressive generation process~\cite
{yu2022orca, kwon2023vllm,lightllm,tgi_hf}. Continuous batching operates at
the iteration level, where completed requests are immediately removed from
the running batch after each iteration to make room for new requests to join.
This allows a new request to be incorporated in just one iteration without
waiting for the entire batch inference to complete. Continuous batching significantly
improves the token generation throughput while minimizing the request queuing
delays. Fig.~\ref{fig:continuous_batching} illustrates this 
batching process used in existing systems~\cite{tgi_hf, kwon2023vllm, lightllm}, where the
decoding and prefill phases interleave as new requests arrive. Upon a request's
arrival, the decoding phase (Dec) is preempted to perform prompt processing,
which involves loading the requested LoRA adapter (Load) and prefilling
(Pre). Once completed, the new requests join the running batch, and the
system combines them together to continue the decoding process.

\subsection{Challenges}
\label{subsec:challenges}

\begin{figure}
    \centering
    \includegraphics[width=0.45\linewidth]{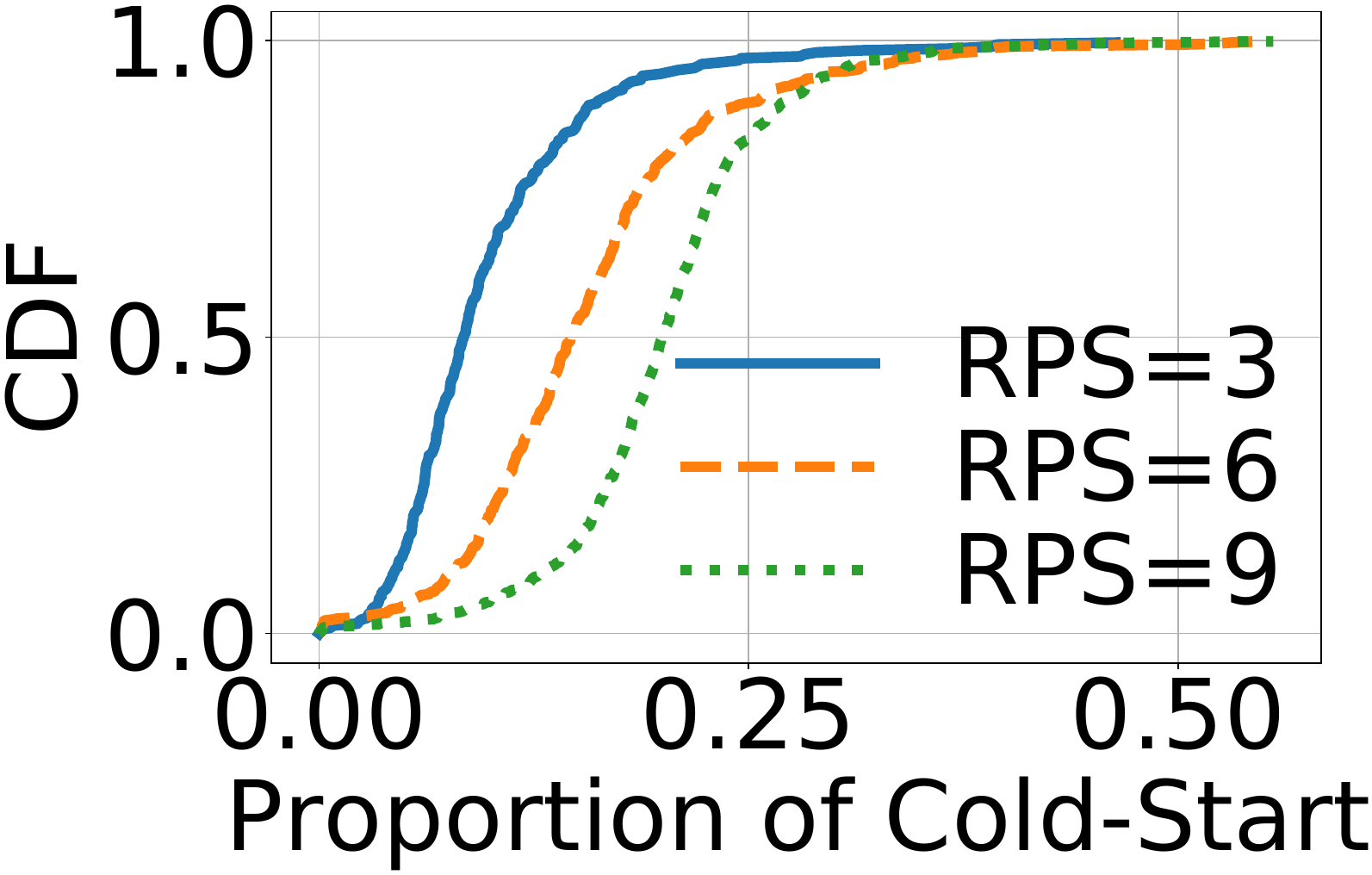}
    \includegraphics[width=0.45\linewidth]{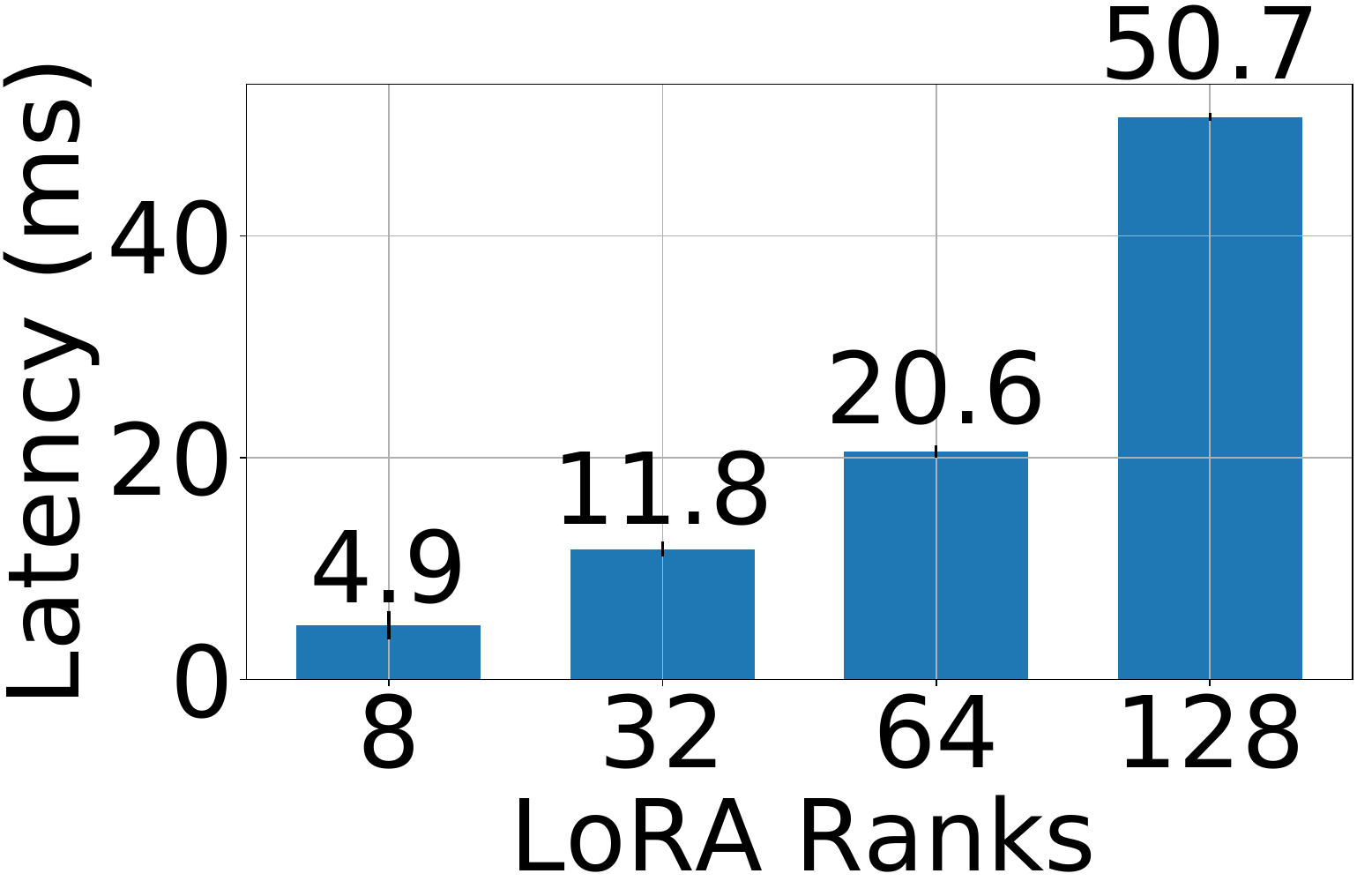}
    \caption{Left: The distribution of cold-start overhead during the entire
    token generation of each request. Right: The cold-start latency of loading a 
    single LoRA adapter of different rank onto GPU. The adapter applies to the $\mathbf{W}_q, \mathbf{W}_k, \mathbf{W}_v$ of a Llama2-7B on an A10 GPU instance.}
    \label{fig:coldstart_portion_loading}
    \vspace{-.15in}
\end{figure}

However, simply enabling LLM-multiplexing and continuous batching is
insufficient to achieve optimal performance for multi-tenant LoRA serving,
as it results in two challenges.


\PHB{C1: High cold-start overhead.}
To save GPU memory, existing systems only cache the base LLM on GPU while
keeping all its LoRA adapters in host memory~\cite
{sheng2023slora,chen2023punica}. When a new request arrives, the system
fetches the corresponding adapter from the host to the GPU, leading to
an \emph{adapter loading} phase that must complete before the prefill phase
begins (Fig.~\ref{fig:continuous_batching}). This results in a severe \emph
{cold-start} problem, where loading an adapter from the host to a GPU can
take between a few to tens of milliseconds, depending on the adapter size
(Fig.~\ref{fig:coldstart_portion_loading}-Right). Cold-start degrades the
service responsiveness, measured by time-to-first-token~\cite
{databricks_metrics, sheng2023slora}. Moreover, under continuous batching,
each time a new request arrives, the decoding phase of in-flight requests
is blocked until the new arrival's prefill phase completes (Fig.~\ref
{fig:continuous_batching}). As new requests keep arriving, their cold-start
overhead \emph{cumulatively delays} the token generation of an in-flight
request (as shown in Fig.~\ref{fig:continuous_batching}, where R1 experiences
two cold-starts due to the arrivals of R2 and R3). We empirically validate
this issue by multiplexing a Llama2-7B model with a group of 512 LoRA
adapters (rank=64). These adapters have skewed popularity (Fig.~\ref
{fig:skewed_lora_dist}) following the Microsoft Azure Function
(MAF) trace~\cite{serverless_in_the_wild}. We configured Poisson request
arrivals with various aggregate loads. Fig.~\ref
{fig:coldstart_portion_loading}-Left shows the proportion distribution of the
cold-start overhead, which, on average, accounts for 10\%, 16\%, and 20\% of
the entire request serving time when the aggregate load is 3, 6, and 9
requests per second, respectively.

To avoid cold-start, a simple approach is to pre-cache all LoRA models in GPU.
However, this approach is expensive: a single rank-64 adapter that
adapts three attention weights $\mathbf{W}_Q, \mathbf{W}_K, \mathbf{W}_V$ of
a Llama2-7B model~\cite{llama2} demands approximately 100~MiB, equivalent to
the size of a KV cache of 200 tokens. S-LoRA~\cite{sheng2023slora} suggests
using predictive pre-fetching, yet without providing details. Given that
inference requests to individual models are highly bursty~\cite
{shepherd, clockwork}, frequent mispredictions and cold-starts are expected.
Punica~\cite{chen2023punica} uses asynchronous loading to avoid blocking
subsequent decoding iterations. However, new requests still need to undergo
the adapter loading phase, leading to the extended time-to-first-token~\cite
{chen2023punica}. 

\begin{figure}[t]
  \centering
  \includegraphics[width=0.9\linewidth]{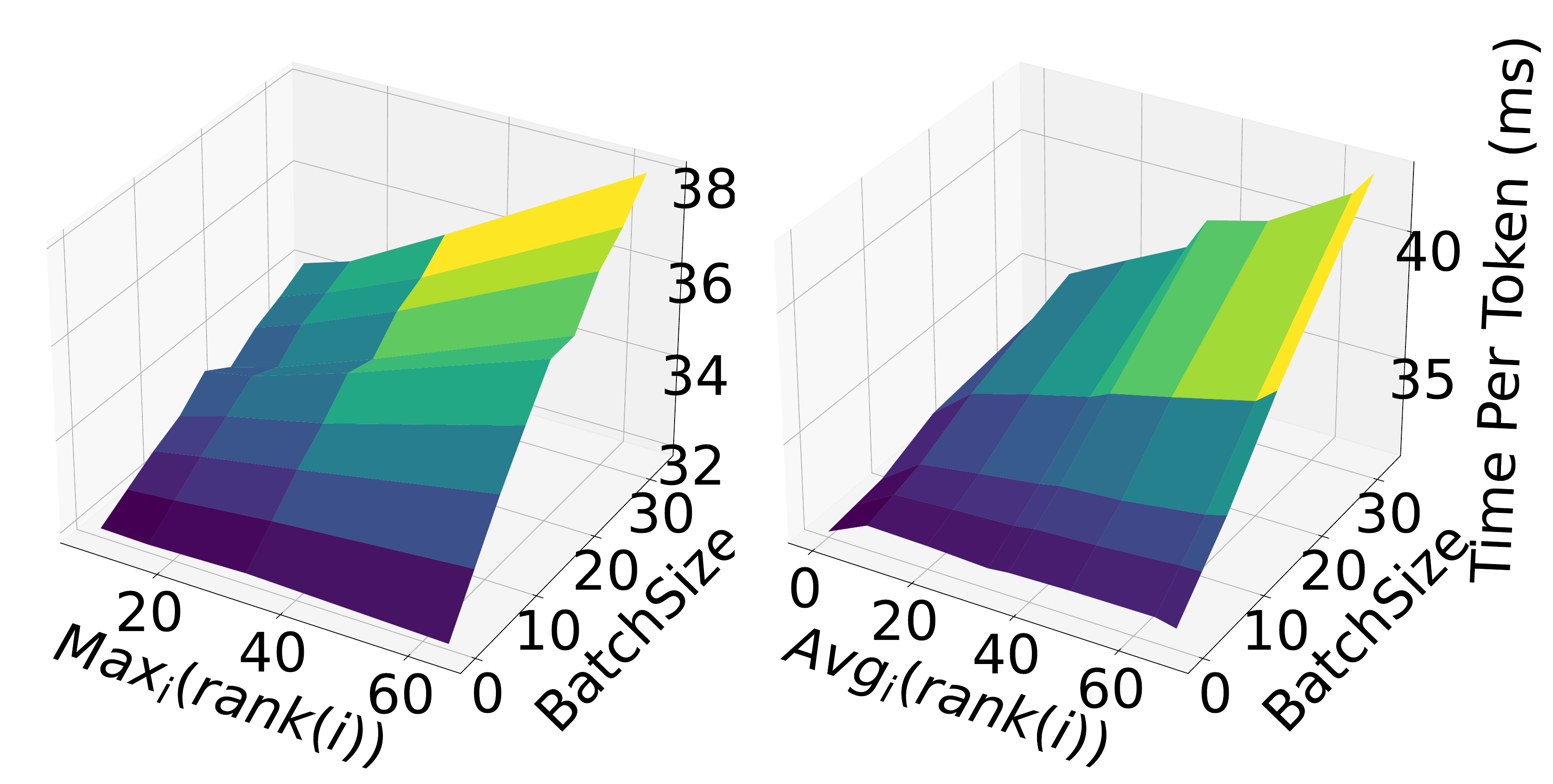}
  \caption{The varying decoding latency of batching heterogeneous LoRA adapters. 
  Left: The performance of Punica's \texttt{BGMV}~\cite{chen2023punica} is determined by the batch size and the \textit{maximum rank}. 
  Right: The performance of S-LoRA's \texttt{MBGMV}~\cite{sheng2023slora}  depends on the batch size and the \textit{average rank} in the batch.}
\label{fig:slora_punica_cost_model}
\vspace{-.15in}
\end{figure}

\PHB{C2: Request scheduling for heterogeneous LoRA serving.}
In multi-tenant LoRA serving, users often request to use heterogeneous LoRA
adapters with varying ranks~\cite{sheng2023slora}. These heterogeneous adapters can be batched
together to multiplex one base LLM using specialized kernel implementations,
such as the Batched Gather
Matrix-Vector Multiplication (\texttt{BGMV}) kernel in Punica~\cite{chen2023punica} or the Multi-size Batched Gather Matrix-Vector Multiplication (\texttt{MBGMV}) kernel
in S-LoRA~\cite{sheng2023slora}. Specifically, when batching a set of
heterogeneous LoRA adapters, \texttt{BGMV} pads adapters of smaller ranks to the
highest rank to perform batch operations, while \texttt{MBGMV} does not use padding~\cite{sheng2023slora}.
As a result, \texttt{BGMV}'s
performance is determined by the maximum rank in the batch, whereas \texttt{MBGMV}'s
performance depends on the average rank. We measure the decoding latency of
batch serving heterogeneous LoRA adapters using these two kernels with
various batch configurations, and the results are depicted in Fig.~\ref
{fig:slora_punica_cost_model}. We observe significant performance variations
when batching different sets of heterogeneous adapters.
This highlights the need for \emph{intelligent request scheduling} that takes
into account the rank heterogeneity and the batching performance of a
specific kernel implementation.

\begin{figure}[tbp]
    \centering
    \includegraphics[width=0.95\linewidth]{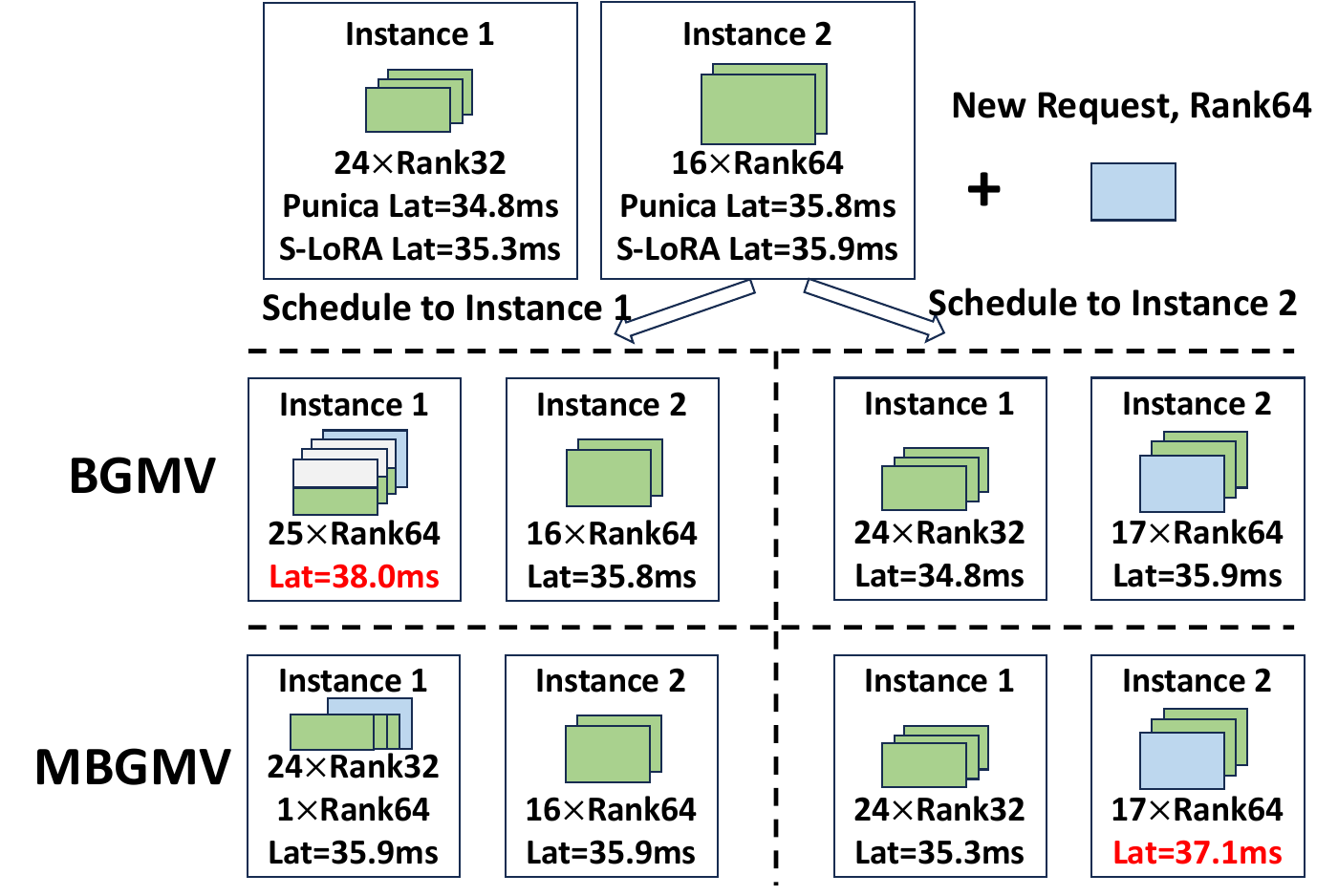}
    \caption{An example of rank-aware LoRA scheduling with a decoding latency SLO of 36 ms. 
    With Punica's \texttt{BGMV}, scheduling the new request to Instance 2 meets the SLO; with
    S-LoRA's \texttt{MBGMV}, scheduling it to Instance 1 preserves the SLO.}
    \label{fig:schedulecase}
    \vspace{-.15in}
\end{figure}

To illustrate this point, we refer to a toy example shown in Fig.~\ref{fig:schedulecase}. In this example, Instance 1 is handling 24
requests with LoRA rank=32, while Instance 2 is running 16 requests with
rank=64. Using Punica's \texttt{BGMV} kernel, the decoding latencies for Instances 1 and 2 are 34.8
ms and 35.8 ms, respectively. With S-LoRA's \texttt{MBGMV}, the latencies are 35.3 ms for
Instance 1 and 35.9 ms for Instance 2. Assume a decoding latency SLO of 36
ms, and we need to determine the optimal schedule for a new incoming request
with rank=64. With the \texttt{BGMV} kernel, assigning this new request to Instance 2 would
meet the SLO, while sending it to Instance 1 would increase the maximum
rank of the batched requests to 64, resulting in an SLO violation due to the
processing of 25 higher-rank requests on Instance~1. Things become different when it
comes to S-LoRA's \texttt{MBGMV} kernel, as the latency is proportional to the total LoRA ranks within a
batch. Since Instance 2 already has a higher sum of batch ranks, its latency is
higher than that of Instance 1. Therefore, scheduling the new request to Instance 1
preserves the SLO, while routing it to Instance 2 would lead to an SLO
violation.

Despite the significant impact of request scheduling, existing LoRA serving
systems~\cite{chen2023punica,sheng2023slora} provide no optimization to it,
resulting in significant delays that violate SLOs (\S\ref
{sec:eval_scheduler}).

\section{\SystemName Overview}
\label{sec:overview}

\begin{figure}
    \centering
    \includegraphics[width=0.85\linewidth]{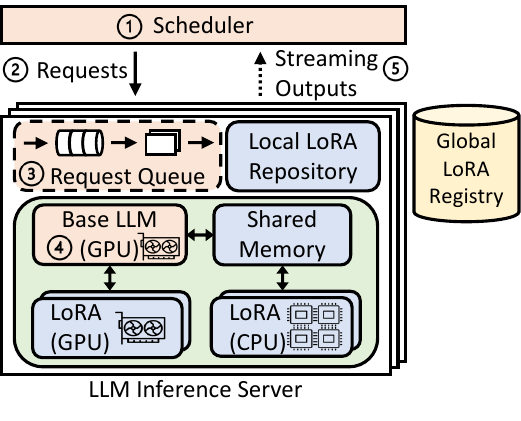}
    \caption{An architecture overview of \SystemName.}
    \label{fig:system_archi}
\end{figure}

In this section, we provide a high-level overview of \SystemName, a LoRA
serving system that efficiently tackles the two challenges mentioned
earlier. \SystemName uses a \emph{CPU-assisted} approach to hide the long
cold-start latency. It uses CPUs to simultaneously execute the requested
LoRA adapter while loading it onto the GPU, effectively overlapping the
adapter loading (cold-start overhead) with the \emph{prefill} computation
(\S\ref{sec:cpu_assisted_lora_serving}). \SystemName also optimizes the
scheduling of inference requests to heterogeneous LoRA adapters using
a \emph{rank-aware} scheduling algorithm, significantly enhancing cluster
performance and SLO compliance (\S\ref{sec:scheduler}). Fig.~\ref
{fig:system_archi} illustrates the system architecture, which consists of a
cluster of LLM inference servers, a scheduler, and a global LoRA registry.

\PHB{LLM inference server.}
Each LLM inference server maintains a long-running service of the base LLM on
the GPU. It also stores a set of heterogeneous LoRA adapters in an in-memory \emph
{local LoRA repository}.
During inference, the server coordinates LoRA computations on the CPU and
GPU to avoid cold-start. Specifically, it adapts the \texttt{BGMV} kernel from~\cite{chen2023punica} to
perform LoRA computation efficiently on the GPU. For CPU-based LoRA
execution, it utilizes three techniques to enhance its efficiency:
asynchronous invocation, shared memory, and profiling-guided
parallelization, which we elaborate in \S\ref{sec:cpu_assisted_lora_serving}.

\PHB{Scheduler.} 
The scheduler receives user requests and routes them to the appropriate servers to
meet the SLOs. To guide the scheduling decision, it uses a performance
model to predict the latency cost by jointly considering the rank
heterogeneity of the serving batch and the underlying kernel implementation,
which we explain in \S\ref{sec:scheduler}.

\PHB{Global LoRA registry.}
The global LoRA registry stores the metadata of all LoRA adapters, 
such as the LoRA ranks, the path to their weights file, etc. 

\PHB{Workflow.}
As illustrated in Fig.~\ref{fig:system_archi}, new requests arrive at the scheduler
(\circled{1}), which uses the rank-aware scheduling algorithm described
in \S\ref{sec:scheduler} to route them to appropriate inference servers
(\circled{2}). Following the continuous batching strategy~\cite
{yu2022orca}, the LLM inference server fetches requests from the request
queue (\circled{3}) and provides generative inference services using the
corresponding LoRA adapters (\circled{4}). New tokens generated by the LLM
are then streamed back to the users (\circled{5}).

\section{CPU-Assisted LoRA Serving}
\label{sec:cpu_assisted_lora_serving}

In this section, we present the design and implementation of CPU-assisted LoRA
serving. We begin by describing LoRA computation on GPU and CPU and
discussing the challenges of efficiently combining the two executions to
address the cold-start problem (\S\ref{sec:gpu_cpu_lora}). We then present
three optimization techniques that address these challenges (\S\ref
{sec:efficient_cpu}).

\subsection{LoRA Computation on GPU and CPU}
\label{sec:gpu_cpu_lora}

A parameter-efficient adapter, LoRA requires lightweight computation and can
run on either GPU or CPU. 

\begin{figure}[t]
\centering
\includegraphics[width=0.9\linewidth]{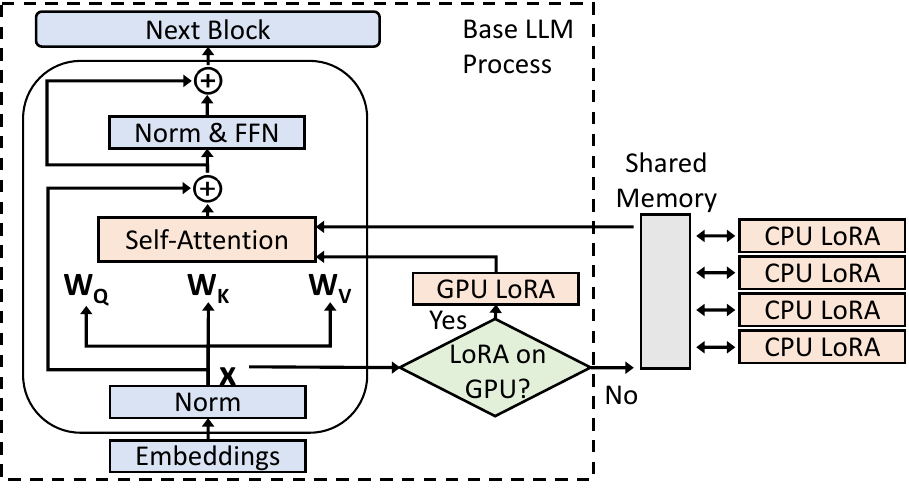}
\caption{Illustration of coordinated LoRA computation on GPU and CPU per transformer block's attention layer.}
\label{fig:lora_flow}
\end{figure}

\PHB{GPU LoRA.}
As the base LLM is ``pinned'' on GPU, running LoRA adapters on the same device
saves the communication overhead and is usually more efficient than running
them on CPU. To maximize the token throughput, LoRA computations
(i.e., $\mathbf{xAB}$ in Eq.~\eqref{eq:xW+xAB}) are batched in each attention
layer during base LLM inference. This can be achieved with a specialized CUDA
operator~\cite{chen2023punica, sheng2023slora}. In \SystemName, we adapt the
Batched Gather Matrix-Vector Multiplication (\texttt{BGMV}) operator~\cite
{chen2023punica}, which parallelizes the LoRA weight gathering and
computation for efficient execution. The LoRA output is then added to the
base output in the self-attention computation, following in Eq.~\eqref
{eq:xW+xAB}. For an efficient implementation, we incorporate the operators of
GPU LoRA computation into the base LLM inference process, as shown in
Fig.~\ref{fig:lora_flow}.

\PHB{CPU LoRA.} 
LoRA computation can also be executed using the CPU, which requires \emph
{layer-wise synchronization} with the base LLM inference running on the GPU.
Specifically, at each attention layer, the base inference process transfers
the input tensor $\mathbf{x}$ in Eq.~\eqref{eq:xW+xAB} from the GPU device
memory to the host memory (Fig.~\ref{fig:lora_flow}). The CPU LoRA process
then performs computation and transfers the result $\mathbf{xAB}$ back to the
GPU device. In the meantime, the base inference process proceeds to compute
$\mathbf{xW}$, which is finally adapted with the received LoRA output
following Eq.~\eqref{eq:xW+xAB}. Although CPU LoRA requires synchronization,
it can start immediately because the LoRA weights are already in memory. We
hence utilize it to address the cold-start problem that arises in GPU LoRA 
(\textbf{C1} in \S\ref{subsec:challenges}).

\PHB{Mitigating GPU cold-start with CPU assistance.}
As illustrated in Fig.~\ref{fig:lora_flow_ite}, when a new request arrives and
the corresponding adapter is not available on the GPU, the server fetches it from
host memory and, in the meantime, starts its \textit
{prefill} computation using the CPU. Once the adapter is fully loaded, the GPU LoRA takes
over, finishing the remaining prefill computation not done by the CPU,
if any, and the subsequent decoding process.
Fig.~\ref{fig:lora_flow} illustrates how CPU and GPU LoRA computations are
coordinated in our design, where we run CPU LoRA adapters as isolated,
concurrent processes for resource/failure isolation and improved performance. 

\PHB{Challenges.}
Though hosting LoRA computation in isolated CPU processes effectively
addresses the \textit{cold-start} problem, it poses three challenges to
system implementation. First, running LoRA in CPU processes requires
layer-wise synchronization between the GPU-based LLM inference to ensure data
validity. Second, frequent triggering of LoRA computation in each attention
layer leads to high invocation overhead, such as inter-process data transfer.
Third, using CPU to compute adaptation can be slow given its limited
parallelization capability, especially when the input prompt is long.


\subsection{Efficient GPU-CPU LoRA Coordination}
\label{sec:efficient_cpu}

In this subsection, we tackle the system challenges mentioned earlier with three
optimization techniques. 

\begin{figure}[t]
\includegraphics[width=0.98\linewidth]
{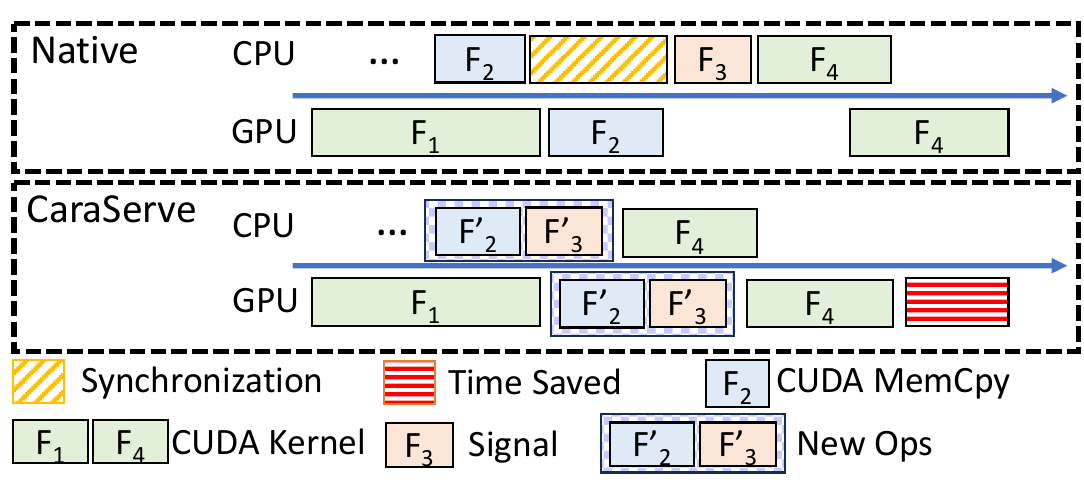}
\caption{Execution timeline of Native LoRA Invocation and LoRA Invocation with \SystemName's operator in base LLM process. CPU LoRA is ignored for simplicity.}
\label{fig:custom_kernel_ops_figure}
\vspace{-.15in}
\end{figure}

\PHB{Sync-free CPU LoRA invocation.}
Most LLM serving systems achieve low latency through asynchronous GPU
computation in PyTorch-like frameworks~\cite
{kwon2023vllm,lightllm,llama2, sheng2023slora, tgi_hf}. However, adapter serving
requires careful coordination between base LLM inference running on GPU
and LoRA invocation running on CPU to ensure correctness and good performance.

In native PyTorch, having the base LLM process invoke CPU LoRA requires
explicit synchronization, which \emph{blocks} subsequent kernels from
launching. To illustrate this problem, we refer to Fig.~\ref
{fig:custom_kernel_ops_figure}-Top, which depicts the native PyTorch
invocation timeline from the base LLM process's perspective.\footnote
{Note that CPU LoRA processes (i.e., CPU calculation for $\textbf{xAB}$) are not depicted in Fig.~\ref
{fig:custom_kernel_ops_figure} because they are identical in both
implementations.} The CUDA kernel $F_1$ computes the input matrix $\mathbf
{x}$. In the meantime, the base LLM process issues $F_2$, a CUDA \texttt
{MemCpy} kernel, to transfer the input matrix to the host memory for CPU
LoRA's access. Once the data transfer completes, the base process uses
a \emph{signaling operator} $F_3$ to notify CPU LoRA processes to compute
$\mathbf{xAB}$. It then launches the next CUDA kernel $F_4$ following $F_1$.
This implementation requires explicit synchronization (shown as a yellow block
with slashes) to ensure that the memory copy ($F_2$) completes before the
signaling ($F_3$). However, this synchronization blocks the subsequent $F_4$
from launching, resulting in significant inference delay and GPU
underutilization.

To address this issue, we introduce a customized operator that eliminates explicit
synchronization by fusing an asynchronous \texttt{MemCpy} kernel with a
signaling kernel. As shown in Fig.~\ref
{fig:custom_kernel_ops_figure}-Bottom, instead of relying on synchronization,
we fuse $F_2$ and $F_3$ into an \emph{asynchronous} CUDA kernel $
[F_2^\prime,F_3^\prime]$, where $F_2^\prime$ performs asynchronous \texttt
{MemCpy} and $F_3^\prime$ asynchronously signals the intended CPU LoRA
processes through shared memory. As a result, the fused kernel $
[F_2^\prime, F_3^\prime]$ can be added to the GPU device queue without
waiting for the completion of $F_1$. Note that data validity is preserved in
this case because CUDA device queue follows a sequential, strict
first-in-first-out execution ordering. Since the new operator requires no
explicit synchronization, subsequent base model kernels, such as $F_4$, can
launch without being blocked, eliminating unnecessary synchronization
overhead. Our experiments in \S\ref{sec:cpulora_mirco} demonstrate that our kernel can
reduce the latency of each prefill iteration by 16\% compared with
PyTorch's native implementation.

\PHB{Shared memory data transfer.} 
Transferring data and signals between the base LLM process and the isolated
CPU LoRA processes requires inter-process communication (IPC). This is a
one-to-N communication involving one base LLM inference process and multiple CPU
LoRA processes. (We explain why multiple CPU LoRA processes later.) We utilize
shared memory for fast inter-process data transfer, eliminating the need for data
copying and serialization (Fig.~\ref{fig:lora_flow}). After the base LLM
process executes our customized operator (see Fig.~\ref
{fig:custom_kernel_ops_figure}), the CPU LoRA processes will soon be signaled
to start reading the input matrix $\mathbf{x}$ from the shared memory and
perform the computation $\mathbf{xAB}$. They then write $\mathbf{xAB}$ back to
the shared memory and notify the LLM inference process to incorporate the adaptation
results (Eq.~\eqref{eq:xW+xAB}). Micro-benchmark evaluations (\S\ref
{sec:cpulora_mirco}) demonstrate that the use of shared memory reduces data
transfer overhead to less than 1 ms (Fig.~\ref{fig:shm_perf}), substantially
outperforming the message passing IPC employed by existing LLM
frameworks~\cite{lightllm}.

\PHB{Profiling-guided LoRA parallelization.} 
Given that the CPU has lower computing power and limited parallelization
capability compared to the GPU, performing LoRA adaptation using a single CPU
is not scalable. Therefore, we propose a profiling-guided parallelization
scheme to accelerate LoRA adaptation using multiple CPU cores. As discussed in 
\S\ref{subsec:llm_inference}, the adaptation computation is $\mathbf{xAB}$, 
where $\mathbf{x} \in \mathbb {R}^{B \times L \times H}$ is the input matrix 
for $B$ requests with $L$ tokens, totaling $B \times L$ tokens. We first profile the
performance achieved by a single core under varying workloads 
(Fig.~\ref{fig:multicpu}-Left) and set
the maximum workload for a single CPU, which is the maximum number of tokens
a CPU core can handle for computation. For example, if one core can handle $c$ tokens,
we allocate $\lceil \frac{L}{c} \rceil$ cores for computing the
adaptation results of each request with weight matrix $\mathbf{W}$. Each core is dedicated 
to an isolated CPU process to avoid interference. Specifically,
the CPU process reads a slice of
$\mathbf{x}$ from the shared memory region, performs the computation, writes
the results back to the shared memory, and notifies the base LLM process accordingly.
Compared to PyTorch's native multi-threading module~\cite
{torch_cpu_thread}, this approach achieves $1.7\times$ speedup when using 8 CPUs
for the same workload (Fig.~\ref{fig:multicpu}-Right).

Putting it altogether, our design, as demonstrated in \S\ref{sec:eval_7b}, can
accelerate the request serving by $1.4\times$ on average.

\section{Rank-Aware Scheduling}
\label{sec:scheduler}

In a multi-tenant LoRA serving system, user requests can trigger the use of
heterogeneous LoRA adapters with varying ranks. As discussed in \S\ref
{subsec:challenges}, the heterogeneity in adapter ranks directly affects the
performance of multi-tenant LoRA serving systems. Therefore, the scheduling
strategy for handling these requests is crucial for enhancing system
efficiency (\textbf{C2}): a sub-optimal strategy can drive the adapter
heterogeneity in a server to a non-ideal setting that slows down token
generation for both new and ongoing requests. To address this, an effective
scheduler needs to be aware of the heterogeneity-performance model, and make
optimal scheduling decisions to achieve high SLO attainment.


\PHB{Performance modeling.} 
The goal of performance modeling is to establish a correlation between rank
heterogeneity in a batch of LoRA requests and its impact on serving
performance. This enables the scheduler to make informed scheduling decisions
to meet SLOs. Under continuous batching (\S\ref
{sec:continuous_batching}), when new requests are routed to a server, the
server's running batch size increases, and the batch's rank heterogeneity
changes as well. To efficiently serve a batch of LoRA requests, existing works~\cite
{chen2023punica,sheng2023slora} provide two CUDA kernels for computing the
adaption $\mathbf{xAB}$: the padding-based \texttt{BGMV} and
padding-free \texttt{MBGMV} (\S\ref{sec:background}). We characterize these
kernels using NVIDIA Nsight Compute~\cite{nsight_compute} and observe that both
kernels consume over $70\%$ of the GPU memory bandwidth, suggesting that
their performance is bounded by the GPU memory bandwidth.


\begin{figure}
    \centering
    \includegraphics[width=0.95\linewidth]{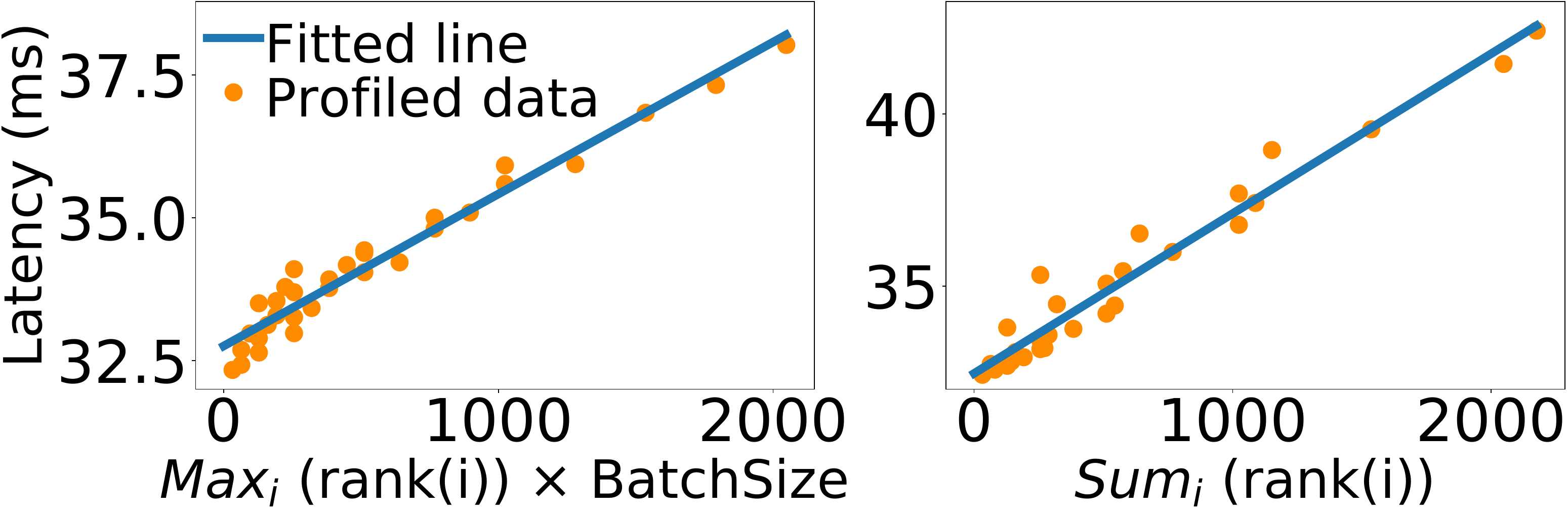}
    \caption{Performance models for \texttt{BGMV} (Left) and \texttt{MBGMV} (Right) kernels. 
    Both linear regression models achieve a high coefficient of determination ($R^2$) of 0.96. }
    \label{fig:linear_costmodel}
    \vspace{-.15in}
\end{figure}

Based on the characterization of kernels, we develop generic performance models to predict the \textit{prefill} and \textit{decoding} latency of a specific batch of heterogeneous adapters.
These models are created through lightweight serving performance profiling, involving varying batch sizes and heterogeneous adapters on a specific GPU.
We present the performance models tailored for both \texttt{BGMV}~\cite{chen2023punica} and \texttt{MBGMV}~\cite{sheng2023slora}. 
For the padding-based \texttt{BGMV} kernel, where lower-ranked LoRAs require padding to match the highest rank for the \texttt{BGMV} operation, we observe that the serving performance of decoding latency is almost linear to the product of batch size and the \textit{maximum} rank encountered in the batch (see Fig.\ref{fig:linear_costmodel}-Left).
On the other hand, S-LoRA's \texttt{MBGMV}~\cite{sheng2023slora} modifies the \texttt{BGMV} kernel to eliminate padding, improving performance with highly heterogeneous LoRA ranks but introducing additional performance overhead for computing homogeneous ranks.
Through profiling, we find that under \texttt{MBGMV}, the serving performance scales linearly with the \textit{sum} of LoRA ranks in a batch of heterogeneous adapters (Fig.~\ref{fig:linear_costmodel}-Right).
Denoting the adapter rank of request $i$ as $rank(i)$, we present performance models for these two kernels on a batch of requests $\mathcal{S}$ as two linear functions with parameters $\alpha$ and $\beta$, inspired by~\cite{orion}:
\begin{equation*}
\begin{array}{cc}
\textsc{Perf}_{\texttt{BGMV}}(\mathcal{S}) = \alpha_B \cdot |\mathcal{S}| \cdot \text{\texttt{Max}}_{i\in \mathcal{S}}rank(i) + \beta_B \\
\textsc{Perf}_{\texttt{MBGMV}}(\mathcal{S}) = \alpha_M \cdot \text{\texttt{Sum}}_{i\in \mathcal{S}}rank(i) + \beta_M
\end{array}
\end{equation*}

As depicted in Fig.~\ref{fig:linear_costmodel}, our linear performance models
accurately fit the profiled data. Both models achieve a high coefficient of
determination ($R^2$) of 0.96, in that $R^2 = 1$ indicates a perfect fit of
the linear model to the data.

\PHB{Scheduling policy.} 
Using the established performance models, we develop a rank-aware
scheduling algorithm (Algo.~\ref{alg:scheduling}) for heterogeneous LoRA requests. Upon receiving a
new request, the scheduler gathers information about ongoing requests from
all available LLM inference servers. The scheduler identifies potential
candidate servers by matching the base LLM, adapter, and GPU memory
availability. If multiple candidates are found, the scheduler calculates a
total cost score for each candidate server based on the performance model.
This cost score measures the impact of the new requests on the performance of
the server's ongoing requests. If serving the new request would cause a 
violation of the SLO, the cost score is assigned a large penalty. The
scheduler then selects the server with the minimum cost score to handle the new
request.
In our evaluation (\S\ref{sec:eval_scheduler}), this rank-aware scheduling
algorithm achieves a high SLO attainment of up to 99\%, substantially
outperforming other baseline strategies. 

\begin{algorithm}[ht]
\caption{Rank-aware Scheduling Policy}
\label{alg:scheduling}
\footnotesize
\SetKwProg{Fn}{Function}{:}{}
\KwIn{Performance models for \textit{Prefill} and \textit{Decoding}: $PrePerf(\cdot)$, $DecPerf(\cdot)$; average response length:  avg\_resp\_len}

\While{ True }
{
    \text{Request} $i$ arrives;\\ 
    candidates $\leftarrow$ available LLM inference servers\\
    \For {instance in candidates} {
        running\_batch, queue = instance.$GetStats$()\\
        cost = $CalcCost$($i$, running\_batch, queue)\\
        requests = len(running\_batch) + len(queue)\\
        instance.total\_cost = cost * requests\\
    }
    best = min(candidates, key=lambda x: x.total\_cost)\\
    best.serve($i$)
}
\Fn{CalcCost{(req, running\_batch, queue)}}{
    exists = running\_batch + queue\\
    \texttt{\# calculate additional prefilling time}\\
    $\Delta_{prefill}$ = $PrePerf$(queue + req) - $PrePerf$(queue)\\
    \texttt{\# calculate additional decoding time per token}\\
    $\Delta_{decode}$ = $DecPerf$(exists + req) - $DecPerf$(exists)\\
    cost\_score = ($\Delta_{prefill}$ / avg\_resp\_len) + $\Delta_{decode}$\\
    \If {$DecPerf$(exists + req) > SLO} {
        cost\_score += penalty\_score
    }
    \Return cost\_score
}
\SetAlgoLined
\end{algorithm}
\vspace{-.15in}
\section{Implementation}

\PHB{LLM inference server.}
We implemented \SystemName's LLM Inference Server on top of LightLLM~\cite
{lightllm}, an LLM serving framework based on PyTorch~\cite
{paszke2019pytorch} and Triton~\cite{tillet2019triton}. Specifically, we
extended its Llama2 inference module to incorporate our LoRA adapters. This
allows for easy integration with different LLMs and other popular LLM
inference frameworks such as vLLM~\cite{kwon2023vllm}. We implemented GPU LoRA
adapters by adapting the \texttt{BGMV} kernels in Punica~\cite
{chen2023punica}. Regarding CPU LoRA, we implemented a custom CUDA kernel
(described in \S\ref{sec:efficient_cpu}) as a PyTorch Extension using \texttt{PyBind11}, and built
CPU LoRA on top of PyTorch. Each CPU LoRA adapter runs as an isolated process,
binding to one CPU core using the \texttt{numactl} command. 
To enable efficient batch inference, we utilize the request queue in LightLLM, which facilitates the continuous batching mechanism~\cite{yu2022orca, kwon2023vllm}.

\PHB{Support model parallelism.} 
We employ tensor parallel techniques~\cite{shoeybi2020megatronlm} to support base LLMs that require multiple GPU devices.
Tensor parallelism involves partitioning a weight matrix into multiple chunks along a specific dimension.
Each GPU device holds only one chunk of the entire weight matrix and performs a portion of the computation in parallel~\cite{alpaserve}.
Tensor parallelism may require communication between the participating GPU devices for output merging.
To enable tensor parallelism for LoRA computation, \SystemName partitions the LoRA adapter weights ($\mathbf{B}$ in Eq.\eqref{eq:xW+xAB}) using the same strategy as that of the base LLMs. It performs the computation and incorporates the adaptation results into the inference intermediates in-place, causing no extra communication overhead.

\PHB{Scheduler \& global LoRA registry. }
In our prototype, we implemented the scheduler using Python Flask. It serves as the frontend that receives requests and routes them to LLM inference servers based on Algo.~\ref{alg:scheduling}. For the global LoRA registry, we utilized SQLite in our prototype.
\newcommand{\Cached}{\textsc{Cached}\xspace}
\newcommand{\OnDmd}{\textsc{OnDmd}\xspace}
\newcommand{\slora}{\textsc{S-LoRA}\xspace}
\newcommand{\mostidle}{\textsc{MostIdle}\xspace}
\newcommand{\firstfit}{\textsc{FirstFit}\xspace}
\newcommand{\random}{\textsc{Random}\xspace}

\section{Evaluation}
\label{sec:eval}

We evaluate 
\SystemName using both synthetic and scaled production workloads~\cite{serverless_in_the_wild} in terms of the LLM inference server's serving efficiency (\S\ref{sec:cpu_assisted_lora_serving}) and the scheduler performance across multiple servers (\S\ref{sec:scheduler}).
Our evaluation highlights include:

\begin{itemize}[topsep=5pt, leftmargin=*]
\vspace{-0.25em}
\item \SystemName achieves efficient multi-tenant LoRA serving on both synthetic and real-world workloads, outperforming strong state-of-the-art baselines, e.g., S-LoRA~\cite{sheng2023slora} (\S\ref{sec:eval_7b}). 

\vspace{-0.5em}
\item \SystemName is compatible with model parallelism to support LLMs that require multiple GPUs (\S\ref{sec:eval_13_70b}).

\vspace{-0.5em}
\item \SystemName's optimizations in CPU LoRA execution are effectively illustrated by various micro-benchmarks (\S\ref{sec:cpulora_mirco}). 

\vspace{-0.5em}
\item \SystemName's scheduler achieves high SLO attainment and improves the performance as a cloud service (\S\ref{sec:eval_scheduler}).
\end{itemize}

\subsection{Experimental Setup}
\label{sec:eval_setup}

\PHB{Model and server configurations.} 
We adopt Llama2~\cite{llama2} models with 7B, 13B and 70B parameters for evaluation (details in Tab.~\ref{tab:model_server_config}), where LoRA adapters are applied to $\mathbf{W}_Q$, $\mathbf{W}_K$, and $\mathbf{W}_V$ (\S\ref{sec:background}) in LLM's attention layers following the standard settings~\cite{qlora, hu2022lora, sheng2023slora}\footnote{ 
Following the setting in~\cite{chen2023punica, sheng2023slora}, we use dummy weights for LoRA models, which do not affect system performance. }.

\begin{table}
\footnotesize
    \centering
    \begin{tabular}{cccc}
    \hline
       Base Model  & Hidden Size & Layers & GPU Config.  \\
    \hline    
       Llama2-7B   &  4096 &  32 &            A10 (24G)  \\
       Llama2-13B  &  5120 &  40 & 2 $\times$ A10 (24G)  \\
       Llama2-70B  &  8192 &  80 & 4 $\times$ A100 (80G) \\
    \hline 
    \end{tabular}
    \caption{Model and GPU configurations.}
    \label{tab:model_server_config}
    \vspace{-.15in}
\end{table}

\PHB{Metrics.} We use the following metrics in evaluation, which are considered essential in user-facing LLM serving~\cite{databricks_metrics, sheng2023slora}.

\begin{itemize}[topsep=5pt, leftmargin=*]
\vspace{-0.5em}
\item \textbf{\emph{Time to first token.}} It measures how quickly users start getting the model's output after entering their prompts. 
Low waiting times for a response are essential in real-time interactions. This metric reflects the time required to process the prompt and then generate the first output token.

\vspace{-0.5em}
\item  \textbf{\emph{Time per token.}} It measures the time on average to generate an output token for each user. This metric corresponds with the perceived "speed" of the model.

\vspace{-0.5em}
\item  \textbf{\emph{Request latency.}} It measures the overall time it takes for the model to generate the full response for a request.
\vspace{-0.5em}
\end{itemize}

\PHB{Baselines.} We consider the following baselines.

\begin{itemize}[topsep=5pt, leftmargin=*]
\vspace{-0.5em}
\item \Cached represents an Oracle method where all required LoRA adapters are pre-cached in unlimited GPU memory. 
It has no adapter loading overhead, thus achieving performance upper bound.

\vspace{-0.5em}
\item \OnDmd loads LoRA adapters on demand. It will suffer from the \textit{cold-start} overhead if the required LoRA adapters are not on GPUs.

\vspace{-0.5em}
\item  \slora~\cite{sheng2023slora} represents a state-of-the-art multi-tenant LoRA serving framework, which is also built on top of LightLLM~\cite{lightllm}.
It loads LoRA adapters on demand and uses an adapted CUDA kernel for GPU LoRA computation.


\end{itemize}
\noindent Note that we equip baselines other than S-LoRA with the \texttt{BGMV} kernel~\cite{chen2023punica} to perform GPU LoRA computation for a fair comparison in the single GPU case.

\PHB{Workloads.} We use both synthetic and scaled production workloads in our evaluation.

\begin{figure}[t]
\includegraphics[width=0.99\linewidth]
{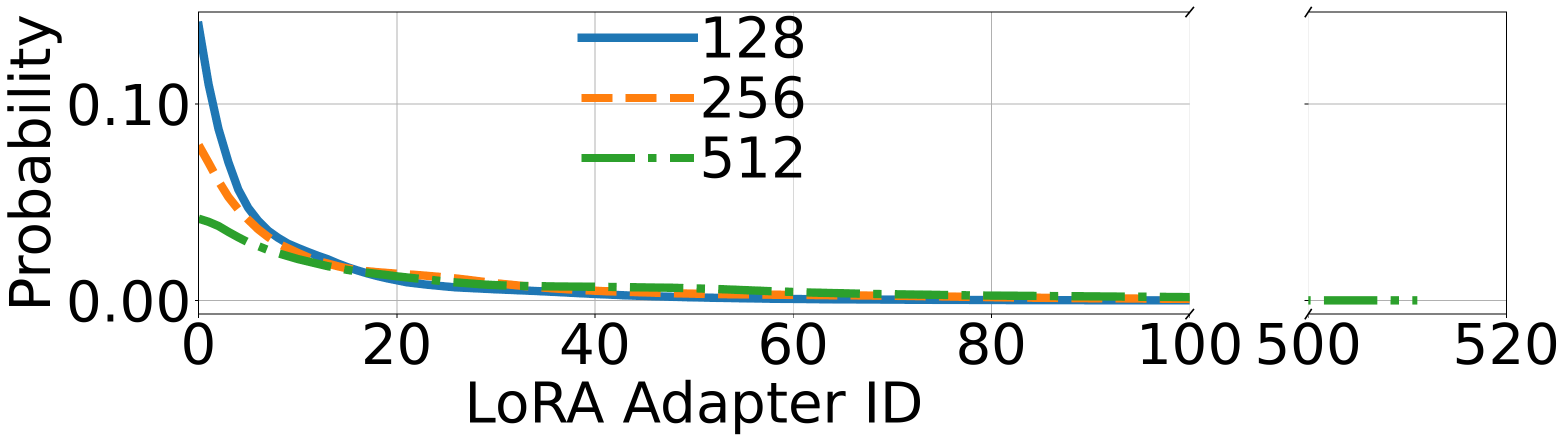}
\caption{LoRA Invocation Probability Mass function. X-axis: ID sorted by invocation probability in descending order.}
\label{fig:skewed_lora_dist}
\vspace{-.1in}
\end{figure}

\begin{itemize}[topsep=5pt, leftmargin=*]
\vspace{-0.5em}
\item \textbf{\emph{Synthetic workload.}} 
The aggregate request traffic to an LLM server follows Poisson processes with varying intensities, widely used in approximating simulated invocations~\cite{shepherd,chen2023punica}. 
Similar to~\cite{chen2023punica}, each request targets a distinct adapter and hence undergoes the adapter loading phase.

\vspace{-0.5em}
\item  \textbf{\emph{Scaled production workload.}} We use the MAF trace~\cite{serverless_in_the_wild} to generate a scaled production workload widely used to emulate model serving workloads~\cite{shepherd, clockwork,INFaaS,spotserve}.
The trace contains invocation patterns of different functions, and we regard each function as one LoRA adapter.
We randomly group the LoRA adapters.
Each LLM inference server hosts a group of adapters and receives the aggregated request traffic from all the LoRA adapters it hosts.
Within a group, adapters have varying probabilities of being invoked, proportional to their invocation frequency in the original trace.
Fig.~\ref{fig:skewed_lora_dist} shows the invocation probability density function. 
\end{itemize}

\noindent For both workloads, we set each request's input prompt and output length according to the Alpaca dataset~\cite{alpaca,kwon2023vllm}, which contains input and output texts of real LLM services.
Like S-LoRA~\cite{sheng2023slora}, we run each workload for 5 minutes.

\begin{figure}[t]
\centering
\includegraphics[width=0.95\linewidth]
{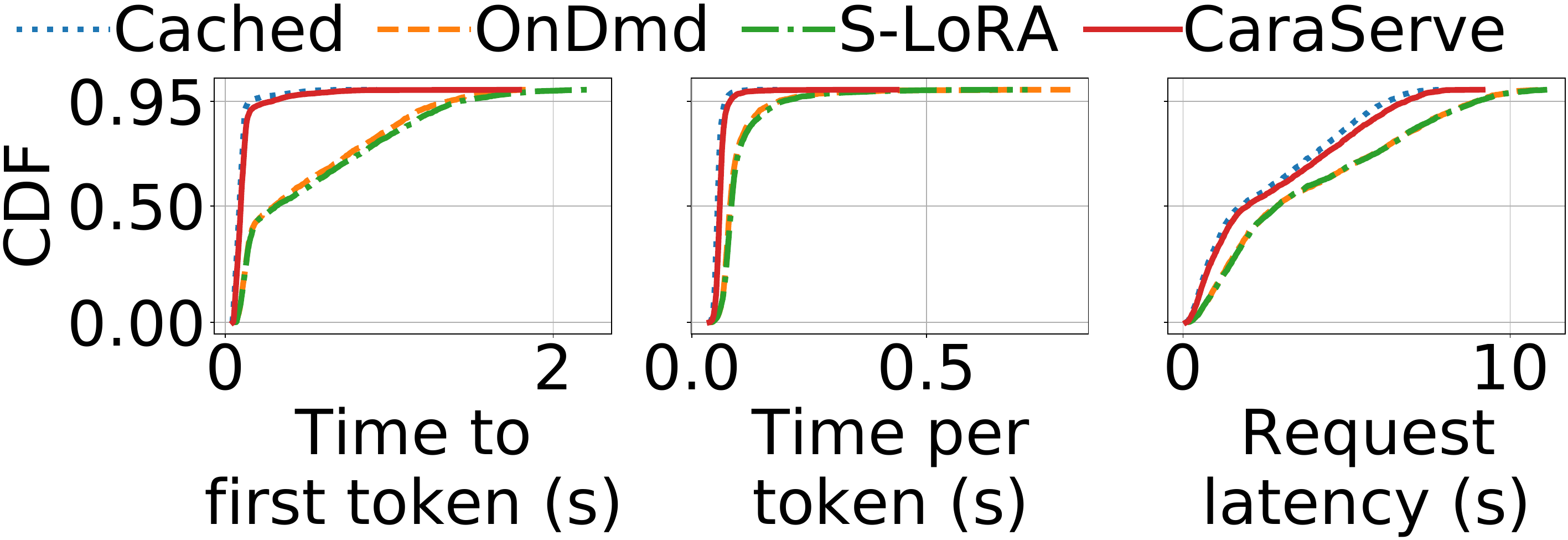}
\caption{End-to-end results with Llama2-7B.}
\label{fig:rank64_rps64_client_metrics}
\vspace{-.1in}
\end{figure}

\begin{figure}[t]
\includegraphics[width=0.99\linewidth]
{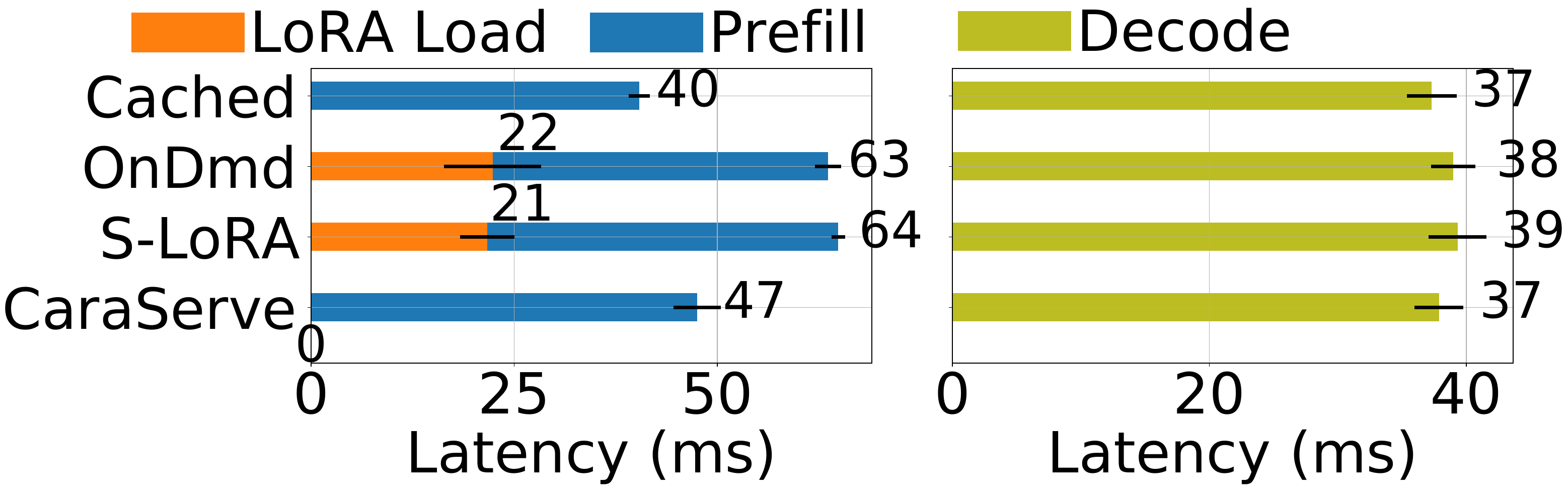}
\caption{Prefill and decoding latency at LLM inference server. \SystemName hides the LoRA adapter loading overhead by overlapping loading and CPU computation.}
\label{fig:rank64_rps64_client_steps}
\vspace{-.15in}
\end{figure}

\begin{figure}[ht]
    \centering
    \includegraphics[width=0.99\linewidth]{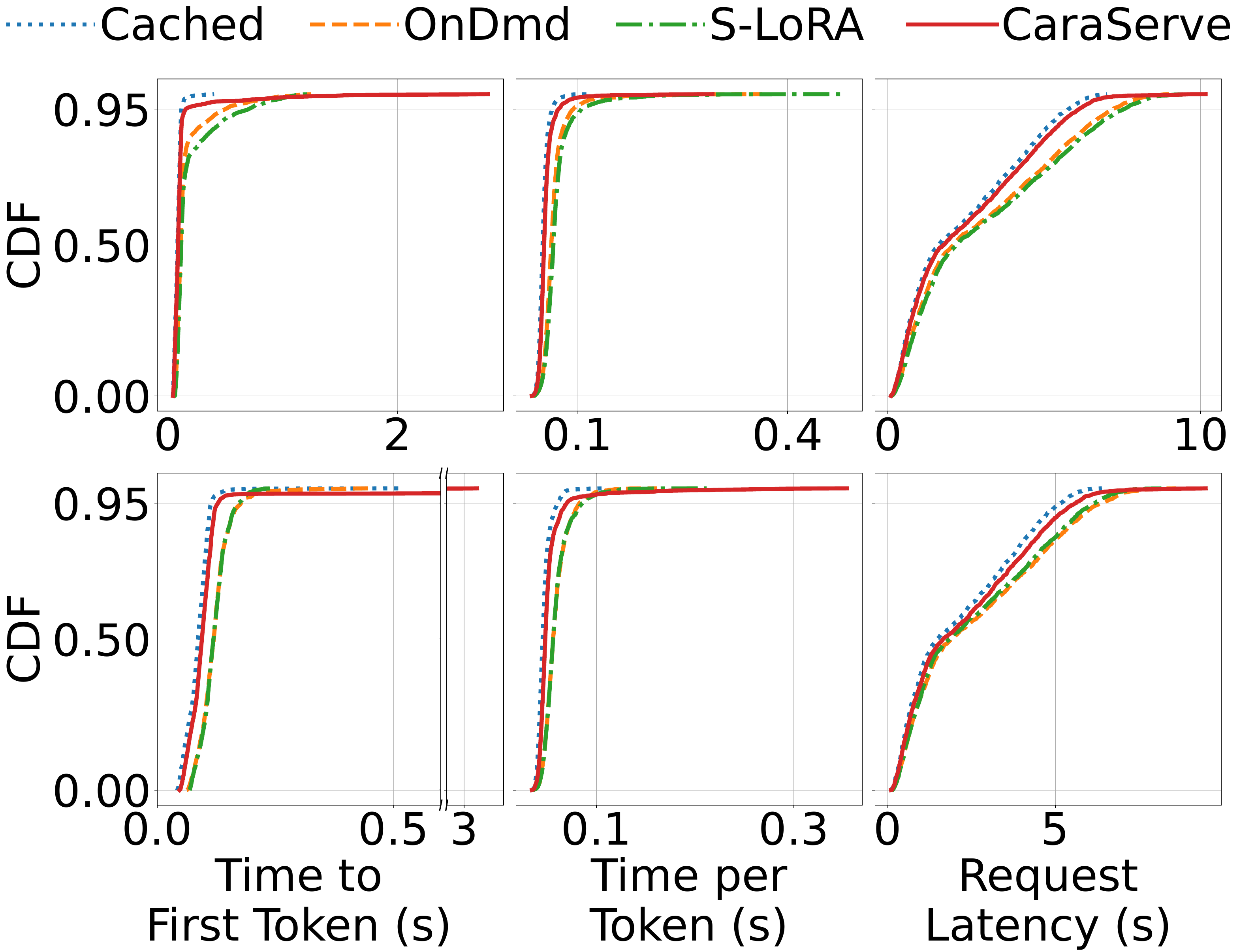}
    \caption{Sensitivity analysis of different Ranks and Traces. Top:
    $RPS=9, rank=32$; Bottom: $RPS=6, rank=64$.}
    \label{fig:sensitivity_analysis}
    \vspace{-.15in}
\end{figure}

\begin{figure}[t]
\includegraphics[width=0.99\linewidth]
{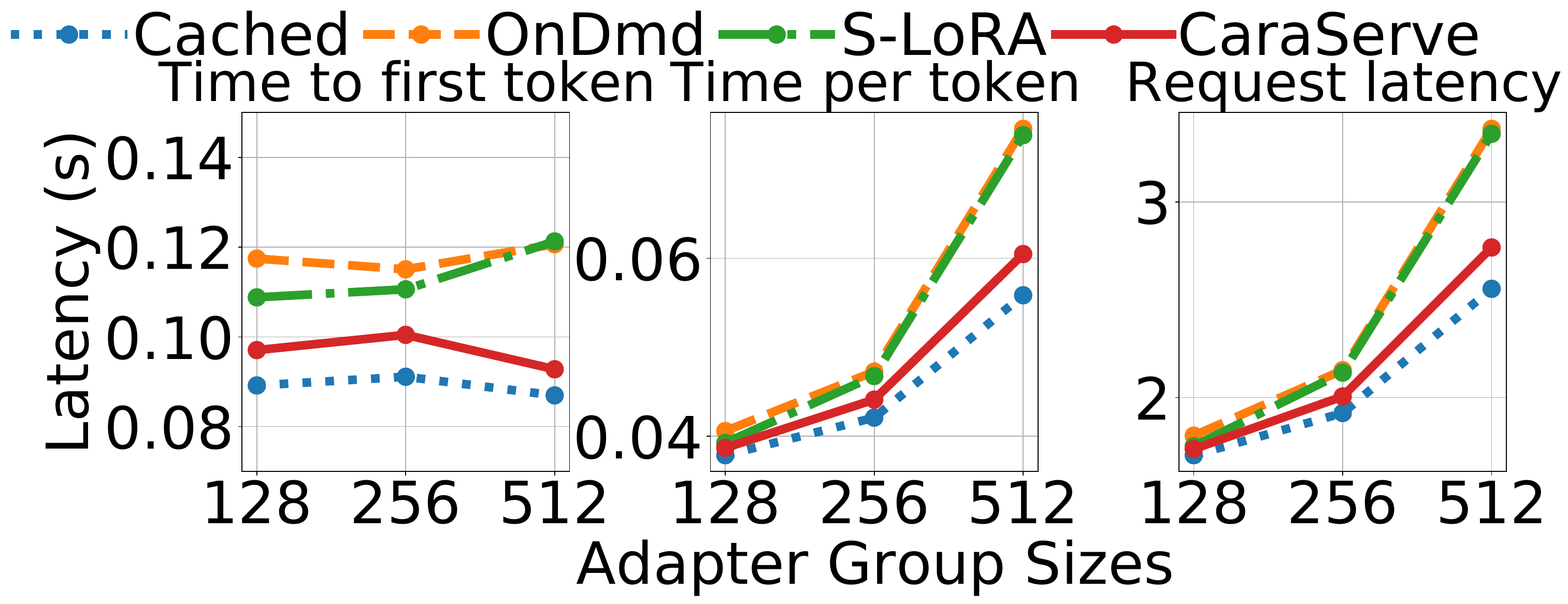}
\vspace{-.1in}
\caption{Baseline performance with varying number of LoRA adapters under MAF workloads.}
\label{fig:maf_loranum}
\vspace{-.1in}
\end{figure}

\begin{figure}[t]
\centering
\includegraphics[width=0.99\linewidth]
{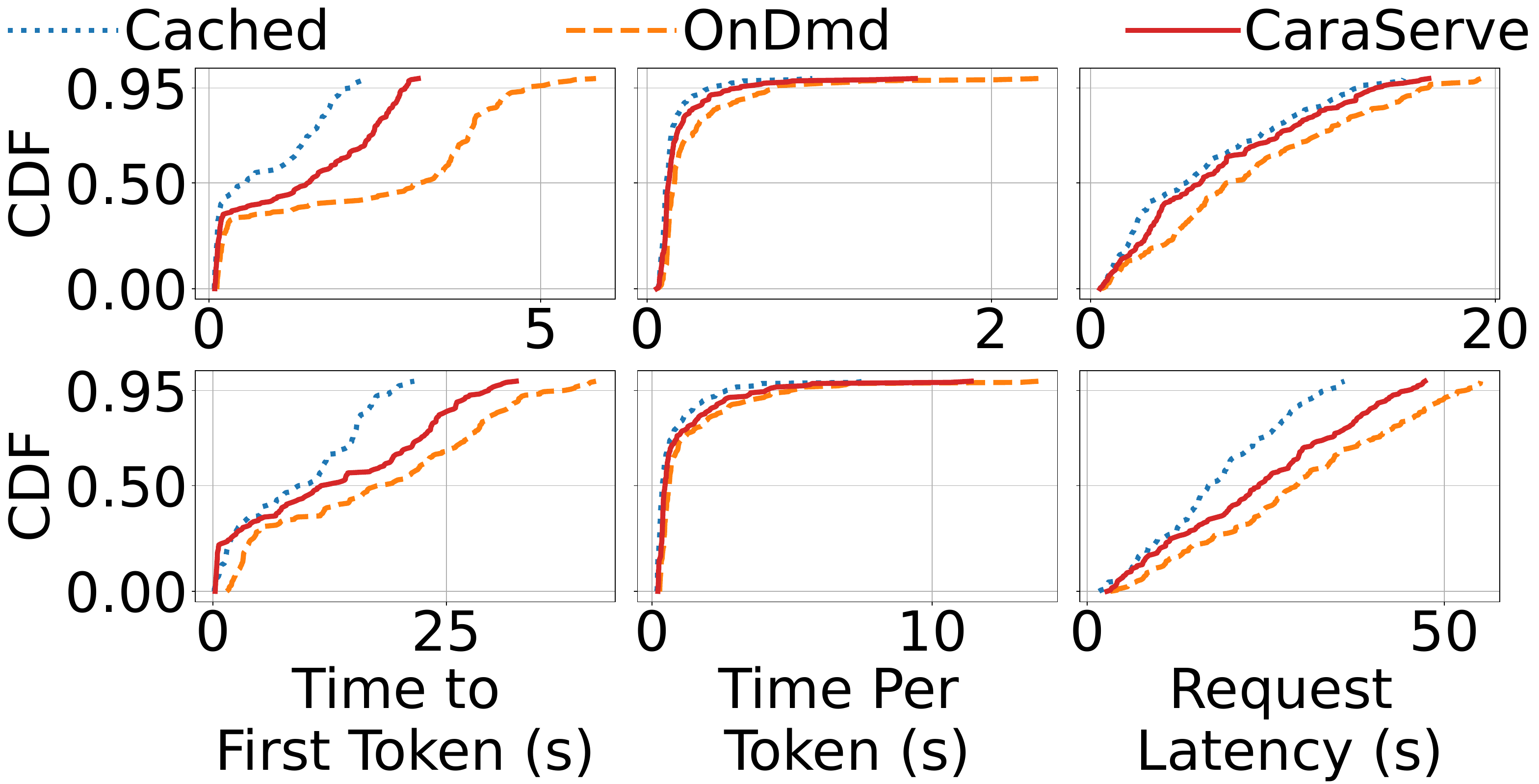}
\vspace{-.1in}
\caption{Evaluation on Llama2-13B (Top) and Llama2-70B (Bottom) models with $RPS=6, rank=64$.}
\label{fig:perf_13b_70b}
\vspace{-.1in}
\end{figure}

\begin{figure}[ht]
    \centering
    \includegraphics[width=0.99\linewidth]{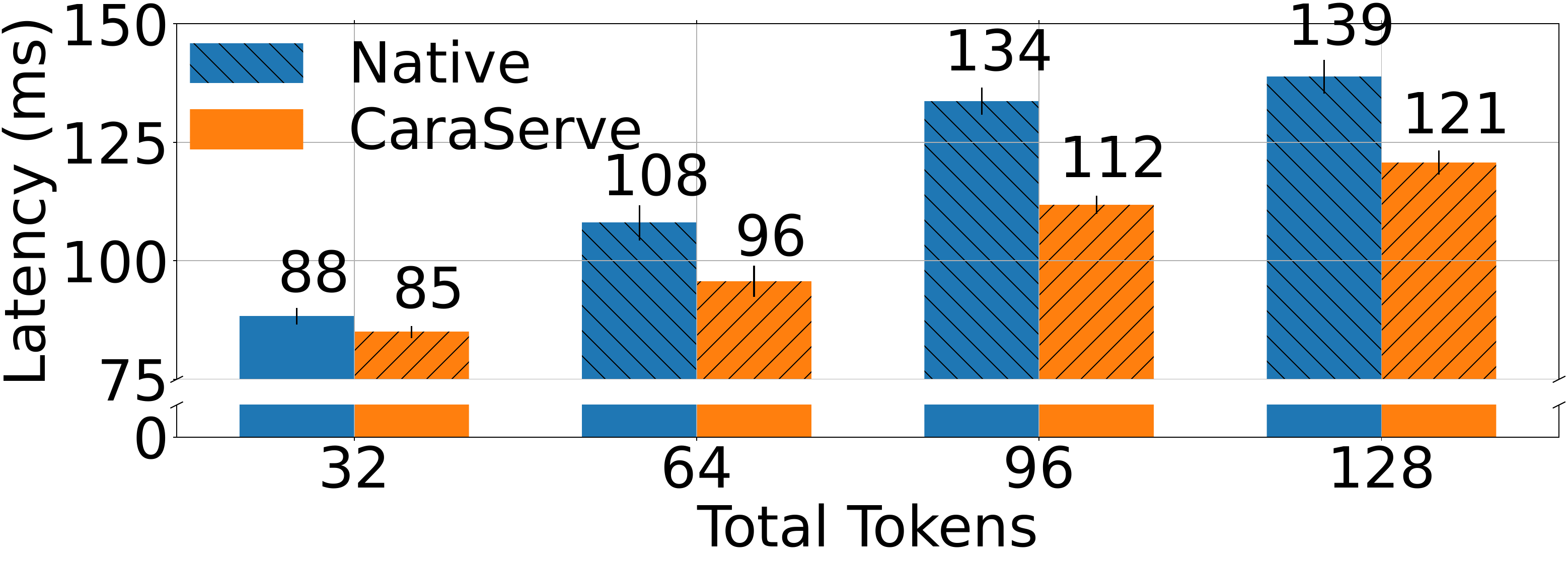}
    \vspace{-.05in}
    \caption{Prefill performance of different kernels on Llama2-7B model. Native: PyTorch default kernels. \SystemName: Implementation with our optimized kernels (\S\ref{sec:efficient_cpu}).}
    \label{fig:custom_kernel_ops}
\end{figure}

\begin{figure}[ht]
    \centering
    \includegraphics[width=0.99\linewidth]{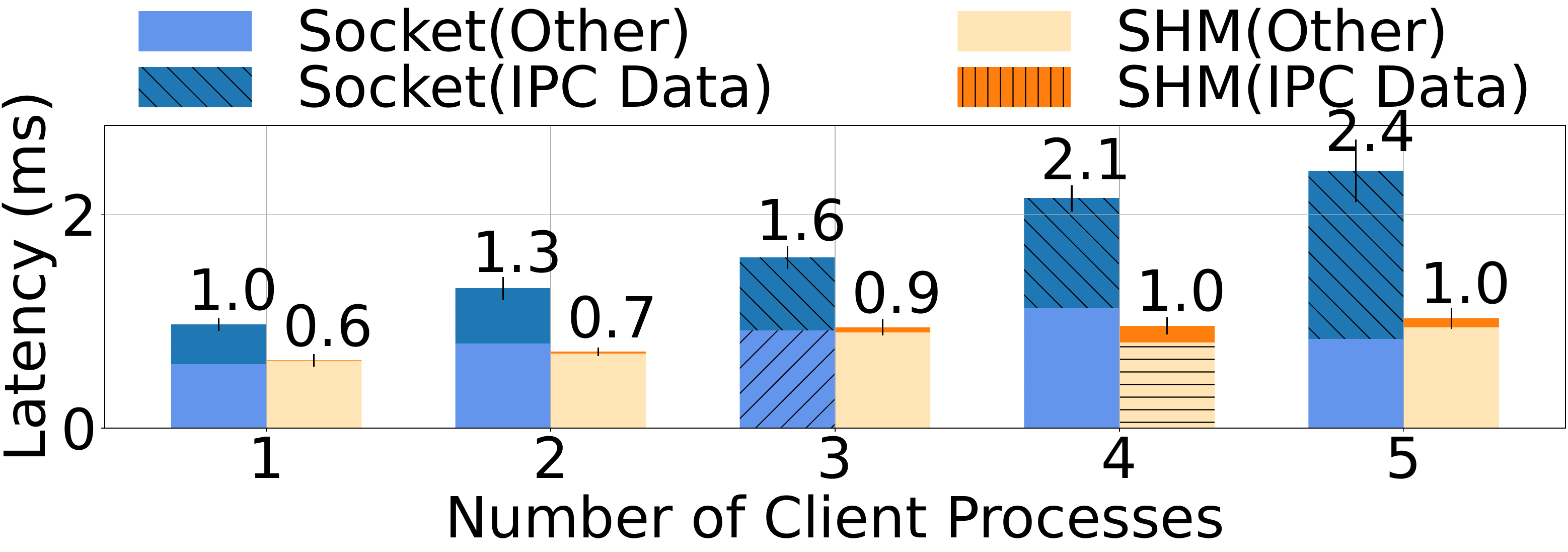}
    \vspace{-.05in}
    \caption{CPU LoRA computation time. Each process receives data of 16 tokens. 
    \textit{Socket}: Domain socket for inter-process communication (IPC). \textit{SHM}: Shared memory for IPC. \textit{IPC Data}: Time for transfering data to another process via IPC. \textit{Other}: Time for all other operations.}
    \label{fig:shm_perf}
    \vspace{-.15in}
\end{figure}

\subsection{End-to-End Performance on a Single GPU}
\label{sec:eval_7b}

We first evaluate \SystemName on the synthetic and scaled production workloads on an A10 GPU serving Llama2-7B. 

\PHB{Synthetic workloads.} 
We generate traces using a Poisson process with an aggregate $RPS=9$ and set the LoRA adapter rank to 64.
We measure the performance of each baseline using the metrics discussed in \S\ref{sec:eval_setup}. 
Fig.~\ref{fig:rank64_rps64_client_metrics} plots the CDFs of time metrics, demonstrating that \SystemName can rival \Cached and outperform \OnDmd/\slora.

Compared to the \Cached baseline, \OnDmd/\slora introduce prohibitively high overhead, increasing time to first token by $412\%/451\%$, time per token by $71\%/78\%$, and request latency by $50\%/50\%$ on average. 
However, \SystemName rivals the performance of \Cached by introducing tolerable overheads. 
On average, \SystemName reduces the time to the first token latency overhead to $22\%$, time per token overhead to $11\%$, and the end-to-end request latency overhead to $9\%$.
Fig.~\ref{fig:rank64_rps64_client_steps} explains \SystemName's advantage from the LLM inference server's side. 
We can see that the latency of each \textit{decoding} iteration is similar across all baselines, while \OnDmd/\slora have a long \textit{prefill} iteration due to the adapter loading overhead.
On the other hand, \SystemName leverages the CPU-assisted design (\S\ref{sec:cpu_assisted_lora_serving}) to avoid the adapter loading overhead in \textit{prefill} iteration.

\textbf{\emph{Sensitivity Analysis.}} 
Two factors affect the benefits achieved by \SystemName (\S\ref{sec:background}).
The \underline{first} is LoRA rank ---
smaller rank leads to shorter loading latency.
We evaluate each baseline with adapter $rank=32$ and aggregate $RPS=9$.
Fig.~\ref{fig:sensitivity_analysis}-Top shows that 
although smaller LoRA ranks decrease overhead, \OnDmd/\slora introduces a considerable amount of overhead compared to the \Cached: $88\%/126\%$ for time to first token, $28\%/36\%$ for time per token, and $25\%/31\%$ for request latency on average.  
\SystemName outperforms by introducing minimal overhead: $36\%, 5\%, 6\%$ for the three metrics respectively.
The \underline{second} factor is the workload, which determines the frequency of LoRA loading. 
Higher request traffic results in increasing \textit{prefill} phases and adapter loading (\S\ref{subsec:challenges}).
We evaluate each baseline with a lighter traffic with aggregate $RPS=6$ and the $rank=64$.
Similar to reducing LoRA rank, reducing workload decreases the overheads of \OnDmd/\slora to $42\%/41\%$, $25\%/25\%$, $24\%/20\%$ for the three metrics respectively (Fig.~\ref{fig:sensitivity_analysis}-Bottom). \SystemName maintains superior with minimal overhead: $1\%, 10\%, 9\%$ for the three metrics, respectively.

\PHB{Scaled production workloads.} 
We next evaluate \SystemName on a production workload based on the MAF trace~\cite{serverless_in_the_wild}.
Fig.~\ref{fig:skewed_lora_dist} illustrates the skewed distribution of function popularity.
We evaluate each baseline with an increasing number of LoRAs and their workloads in a single LLM inference server.
More LoRA adapters mean heavier request loads, and each new request is more likely to invoke a new LoRA adapter that needs to be loaded onto GPU on demand (Fig.~\ref{fig:skewed_lora_dist}). 
The average aggregate $RPS$ for $128/256/512$ adapters is $1.5/3.6/7.7$, respectively, scaled from the original trace. 


We measure each request's serving performance using the metrics
defined in \S\ref{sec:eval_setup}.
Fig.~\ref{fig:maf_loranum} presents the results.
When $128$ LoRA adapters are in a single server, the impact of \textit{cold-start} is negligible because the invocation traffic is low, and most new requests do not require adapter loading. 
Compared to \Cached, \OnDmd/\slora/\SystemName increase time to first token by $31\%/22\%/9\%$, time per token by $8\%/3\%/3\%$, and request latency by $6\%/3\%/2\%$ on average.

However, as the number of LoRA adapters increases to $512$, adapter loading introduces prohibitively high overhead, hindering a system from scaling to host a large number of LoRA adapters. 
In comparison to the \Cached baseline, \OnDmd/\slora/\SystemName increase first token latency by $39\%/39\%/7\%$, time per token by $34\%/32\%/7\%$, and request latency by $31\%/31\%/8\%$ on average. 
These results suggest that the \textit{cold-start} issue prevents \OnDmd/\slora from scaling to accommodate a large number of LoRA adapters. 
Nevertheless, \SystemName performs better than its competitors by rivaling the performance of the \Cached baseline.

\subsection{End-to-End Performance on Multi-GPUs}
\label{sec:eval_13_70b}
We evaluate each baseline with Llama2-13B and Llama2-70B with two A10 GPUs and four A100 GPUs, respectively. 
We compare \SystemName with \Cached and \OnDmd since existing works~\cite{chen2023punica,sheng2023slora} have not released their code in multi-GPU settings.
For the Llama2-70B model, we adopt the \texttt{torch.bmm} operator instead of the \texttt{BGMV} kernel from Punica~\cite{chen2023punica}, since it does not support the key/value matrix shape of the Llama2-70B model.
For both models, we use a synthetic Poisson arrival rate with $RPS=6$ and prompts from the Alpaca dataset~\cite{alpaca}. 

Fig.~\ref{fig:perf_13b_70b} plots the CDFs of requests' serving performance regarding the three metrics.
\SystemName gains a much better performance than the on-demand loading methods. 
On average, \SystemName achieves a $20.2\%/18.5\%$ speedup on the end-to-end request latency for Llama2-13B and Llama2-70B models. 
Compared with \OnDmd, \SystemName reduces its cold-start overhead by over $50\%$.

\subsection{Microbenchmark Evaluation}
\label{sec:cpulora_mirco}
In this section, we evaluate \SystemName's optimizations on CPU LoRA computation (\S\ref{sec:efficient_cpu}) at a micro-benchmark level.

\PHB{Sync-free CPU LoRA invocation.}
To analyze the performance of our optimized CPU LoRA invocation kernel, 
we use the Llama2-7B model on one A10 GPU to measure the \textit{prefill} latency of the PyTorch's native implementation and our optimized kernels.
As Fig.~\ref{fig:custom_kernel_ops} shows, our customized kernel performs better than the default PyTorch kernel. 
As the total number of tokens in \textit{prefill} phase increases, \SystemName's kernel gains up to a $16\%$ performance increase.

\PHB{Shared memory data transfer.}
We compare the latency of computing CPU LoRA with different IPC methods: shared memory and UNIX domain socket. 
We measure the time it takes to perform LoRA computation and the data round trip cost. 
We increase the number of receiver processes to represent the increase in the number of CPU LoRA processes (\S\ref{sec:efficient_cpu}).
Fig.~\ref{fig:shm_perf} shows that as the number of receiver processes increases, the domain socket-based approach suffers from linear time increase in initialization and serialization overhead, whereas the shared memory-based approach obtains near-constant performance. 

\begin{figure}[t]
    \centering
    \includegraphics[width=0.95\linewidth]{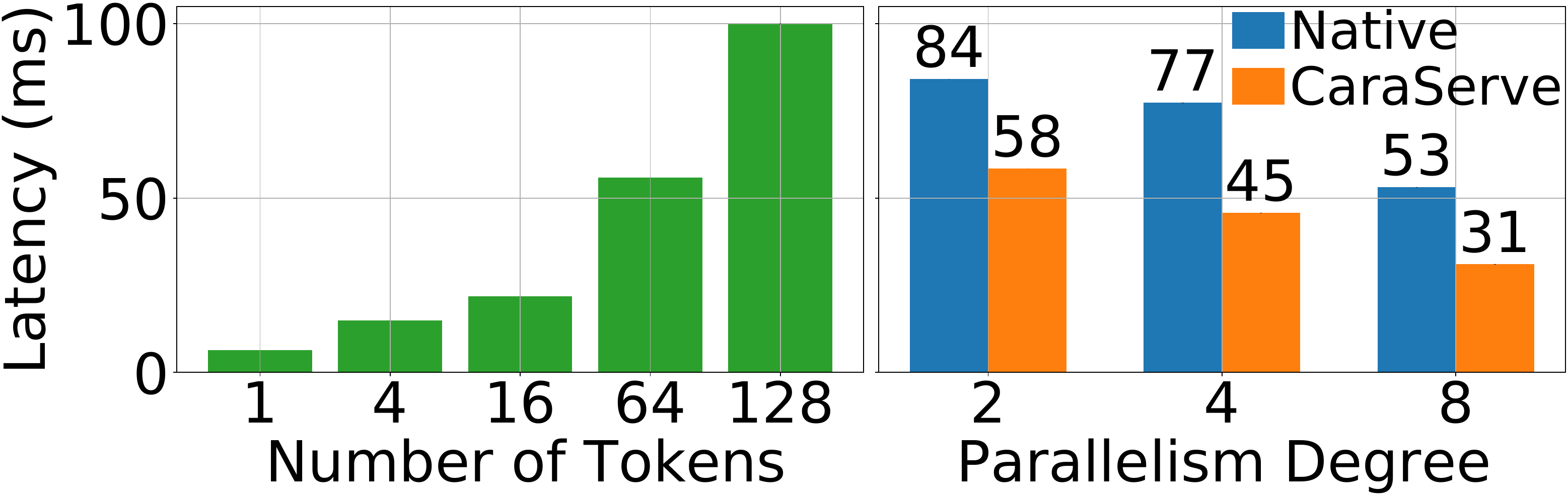}
    \vspace{-.1in}
    \caption{Left: CPU computation time of $\mathbf{xAB}$ in the \textit{prefill} phase for prompts of different length. Right: Comparison of CPU computation time  for 128 tokens with different CPU parallelism. Native: PyTorch native multi-threading~\cite{torch_cpu_thread}.}
    \label{fig:multicpu}
    \vspace{-.1in}
\end{figure}

\PHB{Multi-CPU computation.}
We first profile the LoRA computation performance during a \textit{prefill} phase with a single CPU.
We profile it with different workloads (number of tokens to process).
We run the profiling on a Llama2-7B model with a single A10 card. 
As shown in Fig.~\ref{fig:multicpu}-Left, the CPU has limited parallelism and does not scale to fit high workloads. 
Fig.~\ref{fig:multicpu}-Right illustrates the performance of \textit{prefilling} a prompt of 128 tokens with \SystemName's multi-CPU design (\S\ref{sec:efficient_cpu}) or the native multi-core utilization of PyTorch multi-threading module~\cite{torch_cpu_thread}.
We can see that \SystemName's design achieves up to $1.7\times$ speedup.

\subsection{Scheduler}
\label{sec:eval_scheduler}
In this section, we evaluate the effectiveness of our scheduling policy (\S\ref{sec:scheduler}), which achieves a higher SLO attainment.

\PHB{Baselines.} Upon the arrival of new requests, we consider the following scheduling baselines for comparison:

\begin{itemize}[topsep=5pt, leftmargin=*]
\vspace{-0.5em}
\item \mostidle scheduler selects the inference server that has the least workload.

\vspace{-0.5em}
\item \firstfit scheduler picks a server following the first-fit bin-packing strategy, which is also adopted by Punica~\cite{chen2023punica}.

\vspace{-0.5em}
\item \random scheduler randomly picks an inference server.
\vspace{-0.5em}
\end{itemize}

\PHB{Setup.} Following~\cite{shepherd}, we run experiments in two settings: a large-scale simulation and an 8-instance real-world testbed.

\begin{figure}[t]
    \centering
    \includegraphics[width=0.9\linewidth]{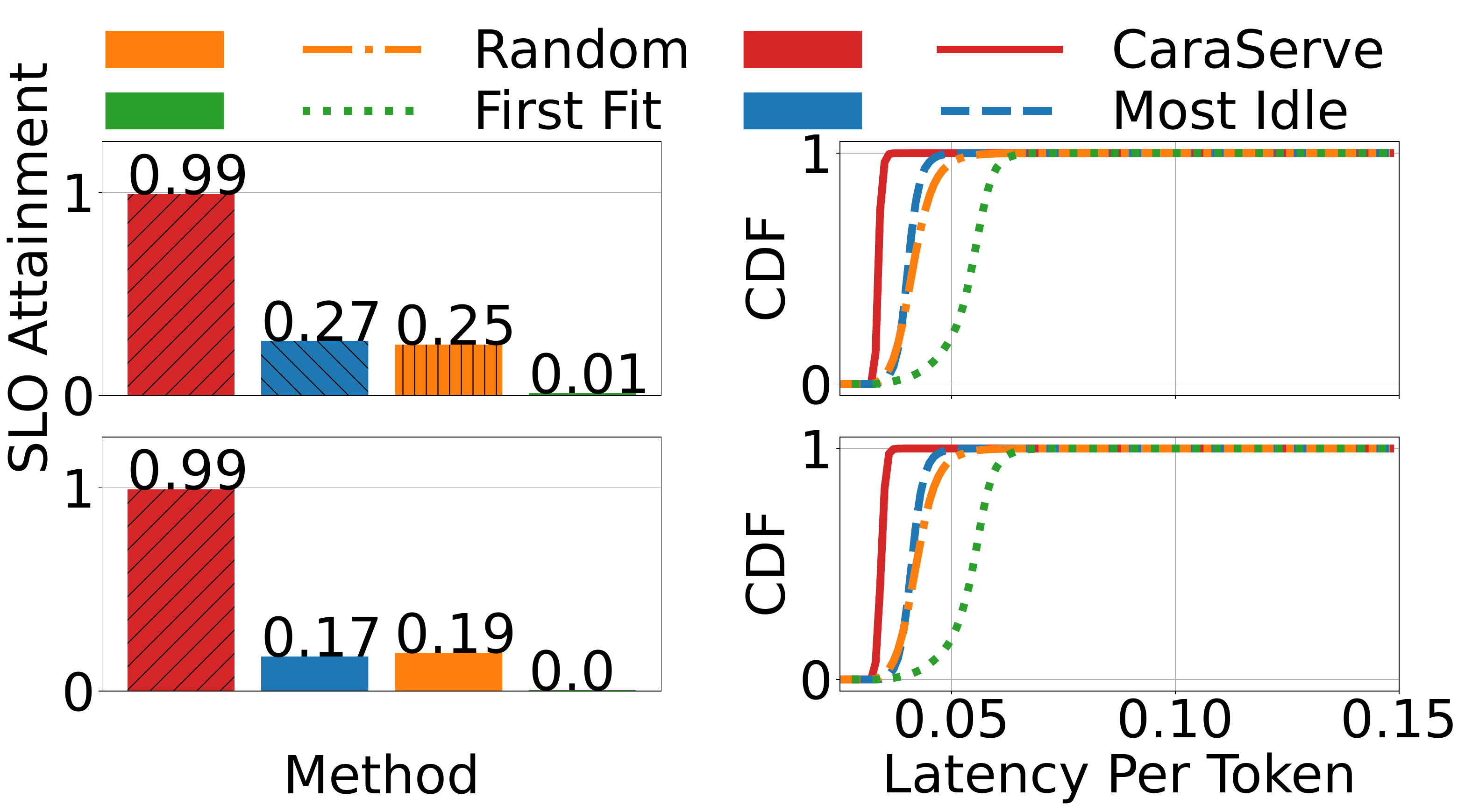}
    \caption{[\textbf{Simulation}] Scheduler performance with S-LoRA's \texttt{MBGMV} and \SystemName's \texttt{BGMV} backend on 60 instances. Top: SLO attainment and time per token CDF with \texttt{MBGMV}. Bottom: Same metrics for \texttt{BGMV}.}
    \label{fig:simulation}
\end{figure}

\PHB{Large-scale simulation.} 
We first evaluate the scheduler's performance through simulation, where we obtain the \textit{prefill} and \textit{decoding} latency of the simulator by profiling. 
We include all 40,000 functions from the MAF trace~\cite{serverless_in_the_wild}, with aggregated $RPS\approx340$, and use 60 simulated servers. 
We set the SLO regarding time per token, as it corresponds to the perceived "speed" of the inference service.
The SLO is set to $1.5\times$ higher than that achieved by the HF-PEFT solution (\S\ref{sec:intro}).
Fig.~\ref{fig:simulation}-Top shows that with S-LoRA's \texttt{MBGMV}, \SystemName's scheduler achieves an SLO attainment of $99\%$ and speeds up the average time per token by $16.1/18.8/36.4\%$ compared to the \mostidle/\random/\firstfit.
Fig.~\ref{fig:simulation}-Bottom shows the performance with Punica's \texttt{BGMV} kernel, which is also adopted in \SystemName (\S\ref{sec:gpu_cpu_lora}). 
Our scheduler has $99\%$ SLO attainment and accelerates time per token up to $36.0\%$.

\begin{figure}[t]
    \centering
    \includegraphics[width=0.9\linewidth]{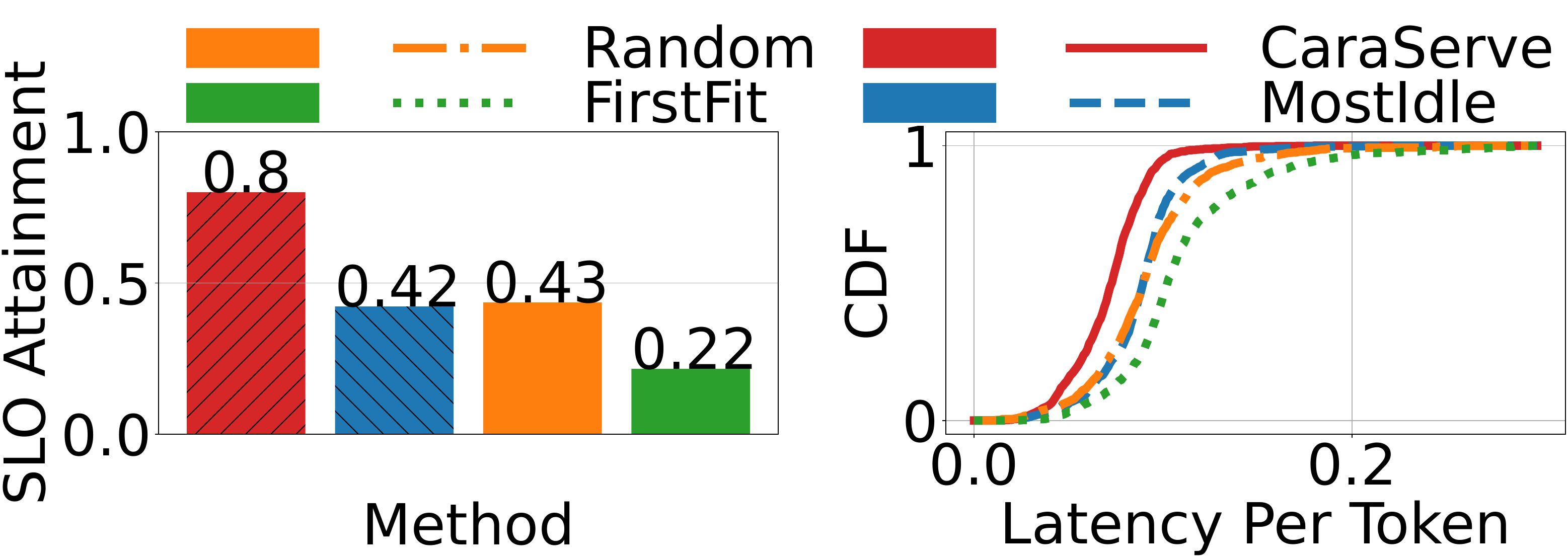}
    \caption{[\textbf{Testbed}] Scheduler performance on 8 instances (\texttt{BGMV}). Left: SLO attainment; Right: Time per token CDF.}
    \label{fig:emulation}
    \vspace{-.1in}
\end{figure}

\PHB{Testbed.} 
Next, we evaluate the scheduler in a small-scale testbed, which has 8$\times$A10 GPUs to support 8 Llama2-7B models.
Due to the limited number of available CPUs, we use \Cached (\S\ref{sec:eval_setup}) as the LoRA serving backend, as our CPU-assisted design can rival its performance in various settings (\S\ref{sec:eval_7b}, \S\ref{sec:eval_13_70b}).
We randomly sample 1,200 requests with an aggregated $RPS\approx60$ from the MAF trace~\cite{serverless_in_the_wild}.
The SLO is also set regarding time per token, which is $1.5\times$ higher than that achieved by the HF-PEFT solution. 
As illustrated in Fig.~\ref{fig:emulation}, \SystemName outperforms other baselines by achieving the highest SLO attainment of 80\%.

\section{Related Work and Discussion}
\label{sec:discuss}

\PHB{LLM inference.}
Optimizing LLM inference is the target of recent studies. 
Orca~\cite{yu2022orca} proposed iteration-level continuous batching to improve the throughput of LLM serving. 
Further, vLLM~\cite{kwon2023vllm} addressed the issue of the GPU memory fragmentation resulting from LLM's KV Cache, improving serving throughput by high GPU efficiency.
FlexGen~\cite{sheng2023flexgen} supported LLMs with limited GPU memory, maximizing serving throughput by efficiently storing, accessing, and quantizing tensors.
SpotServe~\cite{spotserve} leveraged preemptible GPU instances on clouds to reduce the serving cost.
\SystemName is compatible with these optimizations, has already supported continuous batching, and has employed optimized GPU memory management mechanism~\cite{lightllm} to mitigate fragmentation.

\PHB{Multi-tenant LoRA serving.}
Multi-tenant LoRA serving has recently gained attention in the research community.
Punica~\cite{chen2023punica} and S-LoRA~\cite{sheng2023slora} are pioneering works targeting multi-tenant LoRA serving.
They have designed optimized CUDA kernels for GPU LoRA computation and leveraged existing GPU memory management mechanisms~\cite{lightllm,kwon2023vllm} to minimize memory fragmentation. 
These designs are portable to \SystemName.
However, they overlooked the challenges of \textit{cold-start} and \textit{heterogeneity-aware scheduler} (\S\ref{sec:background}).
Besides, PetS~\cite{PetS} proposed a unified framework to serve adapters of different types with LLMs. 
However, it only considered the discriminative language models, which lack an iterative decoding process and continuous batching.

\PHB{Multi-model inference serving.}
A series of works have developed systems for multi-model inference serving, including Clipper~\cite{crankshaw2017clipper}, MArk~\cite{zhang2019mark}, Nexus~\cite{shen2019nexus}, INFaaS~\cite{INFaaS}, Clockwork~\cite{clockwork}, Shepherd~\cite{shepherd}, and AlpaServe~\cite{alpaserve}.
These works optimize batching, caching, model placement, and cost-efficiency in serving multiple models in a cluster.
However, they are not specially designed to serve generative LLMs and heterogeneous LoRA adapters, leading to optimization gaps.

\PHB{Discussion.} 
\SystemName's design is not limited to a particular framework and supports various LLM types. 
Nevertheless, the scalability of CPU-assisted LoRA serving is limited by the number of available CPUs in the host.
Typically, GPU servers designed for LLM serving have abundant CPU cores and host memory. 
For example, the g5.48xlarge instance provided by AWS has 192 vCPU cores.
Such server configurations are also widely used in production clusters~\cite{weng2023mlaas}, where many GPU instances have one A10 GPU and 128 vCPU cores.
In future work, we plan to leverage resource disaggregation to address the scalability issue. 
\section{Conclusion}
This paper presents \SystemName, a multi-tenant LoRA serving system that is GPU-efficient, cold-start-free, and SLO-aware.
In a nutshell, \SystemName exploits base model multiplexing to serve many LoRA adapters in a batch, coordinates LoRA computation on CPU and GPU to avoid cold-start, and employs a rank-aware scheduler to meet SLOs. 
\SystemName is framework-agnostic and can be easily extended to various LLMs.
Compared to existing systems, \SystemName significantly improves serving efficiency by reducing the request serving latency by up to 50\% and achieves an SLO attainment of 99\%.

\bibliographystyle{plain}
\bibliography{sample}

\end{document}